\documentclass[twocolumn,showpacs,preprintnumbers,nofootinbib,prd,
superscriptaddress,10pt]{revtex4-1}

\usepackage{lmodern}
\usepackage{amsmath,amssymb}
\usepackage[normalem]{ulem}
\usepackage{textcomp}
\usepackage{hyperref}
\usepackage{bm}
\usepackage{graphicx}
\usepackage{psfrag}
\usepackage[usenames,dvipsnames]{xcolor}
\usepackage[utf8]{inputenc}
\usepackage{lipsum}
\usepackage{aas_macros}

\hypersetup{
    bookmarks=true,         
    unicode=false,          
    pdftoolbar=true,        
    pdfmenubar=true,        
    pdffitwindow=false,     
    pdfstartview={FitH},    
	pdftitle={High accuracy PN and NR comparisons involving higher modes for eccentric BBHs
and a dominant mode eccentric IMR model},
    pdfauthor={Abhishek Chattaraj et al.},
    pdfsubject={Gravitational Waves},   
    pdfkeywords={gravitational waves, eccentric compact binaries, general 
		relativity},
    pdfnewwindow=true,      
    colorlinks=true,       
    linkcolor=red,          
    citecolor=Blue,        
    filecolor=magenta,      
    urlcolor=Blue,           
    linktocpage=true
}

\allowdisplaybreaks[4]

\DeclareFontFamily{OT1}{pzc}{}
\DeclareFontShape{OT1}{pzc}{m}{it}{<-> s * [1.10] pzcmi7t}{}
\DeclareMathAlphabet{\mathpzc}{OT1}{pzc}{m}{it}

\newcommand{\bigO}{\mathcal{O}}

\newcommand{\hlm}{\mathpzc{h}_{\ell m}}

\graphicspath{{./figures/}}

\begin{document}

\title{High accuracy post-Newtonian and numerical relativity comparisons involving higher modes for eccentric binary black holes and a dominant mode eccentric inspiral-merger-ringdown model}

\date{\today}

\author{Abhishek Chattaraj}
\email{abhishek.chattaraj.ac@gmail.com}
\affiliation{Department of Physics, Indian Institute of Technology Madras, Chennai 600036, India}
\affiliation{Centre for Strings, Gravitation and Cosmology, Department of Physics, Indian Institute of Technology Madras, Chennai 600036, India}
\affiliation{Department of Physics, University of Florida, PO Box 118440, Gainesville, FL 32611-8440, USA}
\author{Tamal RoyChowdhury}
\email{trc.tamal@gmail.com}
\affiliation{Department of Physics, Indian Institute of Technology Madras, Chennai 600036, India}
\affiliation{Centre for Strings, Gravitation and Cosmology, Department of Physics, Indian Institute of Technology Madras, Chennai 600036, India}
\author{Divyajyoti}
\email{divyajyoti.physics@gmail.com}
\affiliation{Department of Physics, Indian Institute of Technology Madras, Chennai 600036, India}
\affiliation{Centre for Strings, Gravitation and Cosmology, Department of Physics, Indian Institute of Technology Madras, Chennai 600036, India}
\author{Chandra Kant Mishra}
\email{ckm@iitm.ac.in}
\affiliation{Department of Physics, Indian Institute of Technology Madras, Chennai 600036, India}
\affiliation{Centre for Strings, Gravitation and Cosmology, Department of Physics, Indian Institute of Technology Madras, Chennai 600036, India}
\author{Anshu Gupta}
\email{anshusm@gmail.com}
\affiliation{Inter-University Centre for Astronomy and Astrophysics, Post Bag 4, Ganeshkhind, Pune 411 007, India}
\begin{abstract}
  Spherical harmonic modes of gravitational waveforms for inspiraling compact binaries in eccentric orbits from post-Newtonian (PN) theory accurate to third post-Newtonian order, and those extracted from numerical relativity (NR) simulations for binary black holes (BBHs) are compared. We combine results from the two approaches (PN and NR) to construct time-domain hybrid waveforms that describe the complete evolution of BBH mergers through inspiral-merger-ringdown (IMR) stages. These hybrids are then used in constructing a fully analytical dominant mode ($\ell$=2, $|m|$=2) eccentric IMR model. A simple extension to a multi-mode model based on this dominant mode model is also presented. Overlaps with quasi-circular IMR waveform models including the effect of higher modes, maximized over a time- and phase-shift, hint at the importance (mismatches $>1\%$) of including eccentricity in gravitational waveforms when analysing BBHs lighter than $\sim 80 M_{\odot}$, irrespective of the binary's eccentricity (as it enters the LIGO bands), or mass-ratio. Combined impact of eccentricity and higher modes seems to become more apparent through smaller overlaps with increasing inclination angles and mass ratios. Additionally, we show that the state-of-the-art quasi-circular models including the effect of higher modes will not be adequate in extracting source properties for signals with initial eccentricities $e_0$ $\gtrsim0.1$.
\end{abstract}

\maketitle
\section{Introduction and summary}
\label{sec:intro}
\vskip 5pt
Since the \emph{first} detection of gravitational waves (GWs) from the merger of two black holes~\cite{GW150914}, the LIGO-Virgo-KAGRA Collaboration has reported over 90 compact binary mergers \cite{GWTC-1, GWTC-2, GWTC-2-extended, LIGOScientific:2021djp}. These include two confirmed mergers of neutron stars~\cite{GW170817, GW190425} and two neutron star-black hole mergers~\cite{2-NS-BH} apart from close to 85 binary black hole (BBH) mergers (see Ref.~\cite{LSC:GWTC-GWOSC} for a complete catalog). Besides providing numerous new insights into the compact binary physics and astrophysics, these observations have proved to be unique probes into binary's ultra-relativistic dynamics~\cite{LIGOScientific:2016lio,LIGOScientific:2018dkp,LIGOScientific:2019fpa,LIGOScientific:2020tif} and have improved our understanding of the underlying astrophysical population of these objects \cite{GWTC-1-pop, GWTC-2-pop, LIGOScientific:2021psn}. However, questions linked with compact binary formation channels largely remain unanswered~\cite{GW150914-astro, GW190521-properties} (see also Ref. \cite{GWTC-2-pop} and references therein). 
A definitive answer to these although can come from the measurements of the orbital eccentricity~\cite{Lower:2018seu}.\\ 

Current template-based search methods make use of quasi-circular templates owing to the expected circularisation of compact binary orbits due to radiation reaction forces~\cite{PhysRev.136.B1224}. However, binaries formed through the dynamical interactions in dense stellar environments or through Kozai-Lidov processes \cite{Kozai62, Lidov62} (if part of stable triples), are likely to be observed in ground-based detectors such as advanced LIGO~\cite{TheLIGOScientific:2014jea} and advanced Virgo~\cite{Acernese:2015gua} with residual eccentricities $e_0\gtrsim0.1$ \cite{Salemi:2019owp}. In fact, the first ever observation of an intermediate mass black hole, GW190521~\cite{GW190521}, hints at it being an eccentric merger of two black holes \cite{GW190521-properties} (see also \cite{Kimball:2020qyd, Romero-Shaw:2021ual, OShea:2021ugg} which also discusses other events with signs of eccentricity apart from the event GW190521). While quasi-circular templates should be able to detect systems with initial eccentricities $e_0 \lesssim 0.1$, binaries with larger eccentricities would require constructing templates including the effect of eccentricity\,\cite{PhysRevD.81.024007, Huerta:2013qb}. Moreover, the presence of even smaller eccentricities ($e_0 \sim$ 0.01-0.05) can induce significant systematic biases in extracting source properties~\cite{GW150914_WF_systematics, Favata:2021vhw}. Furthermore, future ground-based detectors, Cosmic Explorer \cite{McClelland:T1500290-v3, Dwyer:2014fpa, Evans:2016mbw} and Einstein Telescope \cite{Punturo:2010zz, Hild:2010id}, due to their low frequency sensitivities, should frequently observe systems with detectable eccentricities \cite{Lower:2018seu, Tibrewal-etal-2021}.\\

Even though inspiral waveforms from eccentric binary mergers involving non-spinning compact components are sufficiently accurate \,\cite{Mishra:2015bqa, Moore:2016qxz, Tanay:2016zog, Boetzel:2019nfw, Ebersold:2019kdc, koenigsdoerffer-2006, Moore:2019xkm}, waveform models including contributions from merger and ringdown stages compared to quasi-circular versions are less developed. Numerous efforts toward constructing eccentric inspiral-merger-ringdown waveforms useful for data analysis purposes are underway \,\cite{hinder-2018,Huerta:2016rwp, Chen:2020lzc,Setyawati:2021gom}. However, these efforts do not include important physical effects such as spins (aligned and/or precessing) and higher order modes.  Dominant mode ($\ell$=2, $m$=2) models for eccentric BBHs with non-precessing spins were recently developed in Refs.~\cite{Chiaramello:2020ehz, Ramos-Buades:2019uvh}. While one can argue that since most mergers observed so far are consistent with a zero-effective spin\footnote{Only nine of the 44 BBHs reported in the GWTC-2 catalog \cite{GWTC-2} have been identified with a positive effective spin parameter with zero outside the 95\% credible interval.} \cite{GWTC-1-pop, GWTC-2-pop}, models neglecting spin effects can still be useful\footnote{Reference~\cite{OShea:2021ugg} explores correlations between the binary's spins and eccentricity.}~\cite{Huerta:2016rwp}, modeling of higher order modes seems necessary as far as full inspiral-merger-ringdown (IMR) eccentric waveform models are concerned as Ref.~\cite{Rebei:2018lzh} argues and as is also discussed in detail in the current work.  Very recently, eccentric versions of the effective-one-body (EOB) waveforms including higher modes \cite{Ramos-Buades:2021adz, Nagar:2021gss} and an eccentric numerical relativity (NR) surrogate model \cite{Islam:2021mha} appeared online. In the absence of reliable inspiral-merger-ringdown models for eccentric mergers, sub-optimal search methods (with little or no dependence on signal model being searched) are used~\cite{Salemi:2019owp}. While these methods should detect binaries with arbitrary eccentricities, these are sensitive to high mass searches (typically $\gtrsim 50M_{\odot}$) \cite{Salemi:2019owp}, while most observed events have a mass smaller than this limit \cite{GWTC-2-pop, GWTC-2-extended}; see, for instance Fig. 3 of \cite{Divyajyoti:2021uty}. \\ 

\subsection{Summary of the current work}

The current work assesses the impact of neglecting eccentricity as well as of eccentricity-induced corrections to higher modes\footnote{Eccentricity induces sub-dominant modes appearing as oscillating multiples of the mean anomaly (or an equivalent parameter) in each spherical harmonic mode; see, for instance, Eq.~(76) ~\cite{Boetzel:2019nfw}. We refer to these contributions as ``eccentricity-induced corrections to higher modes" throughout the paper.} on detection and parameter estimation of GWs from BBHs in eccentric orbits. The inadequacy of quasi-circular templates in extracting an eccentric IMR signal including higher modes is demonstrated through (simple) mismatch calculations (see Fig.~\ref{fig:match}) using state-of-the-art quasi-circular waveform families including higher modes \textsc{SEOBNRv4HM}~\cite{Cotesta:2020qhw} and \textsc{IMRPhenomXHM}~\cite{Garcia-Quiros:2020qpx}. These mismatch plots indicate the need for including eccentricity in gravitational waveforms for analysing BBH systems with masses below $80M_{\odot}$, irrespective of eccentricity or mass ratio. Additionally, as should be clear from middle and right top/bottom panels of Fig.~\ref{fig:match}, with increasing orbit's inclination with respect to our line of sight, the combined impact of eccentricity and eccentricity-induced corrections to higher modes becomes more apparent.\\

Further, Sec.~\ref{sec:pe_systematics} discusses waveform systematics that may be induced due to the absence of eccentricity in recovery waveforms in a parameter estimation study. This is illustrated through an injection analysis presented in Fig.~\ref{fig:pe_systematics} attempting recovery of an eccentric signal with quasi-circular waveforms (see Sec.~\ref{sec:pe_systematics} for details). Both the injections and the recovery templates include the same set of modes to avoid biases due to additional modes in target/template waveforms.  Figure \ref{fig:pe_systematics} also shows recovery of a circular injection as a reference. Both circular and eccentric injections correspond to a BBH system of total mass 40$M_{\odot}$, and the eccentric simulations have an orbital eccentricity of $e_0\sim0.1$ at
20 Hz. Non-recovery of the injected value of the chirp mass for the eccentric case can be interpreted as the bias induced due to the neglect of eccentricity and associated higher modes.\\

\begin{figure*}[t!]
    \centering
    
    \includegraphics[width=0.46\textwidth]{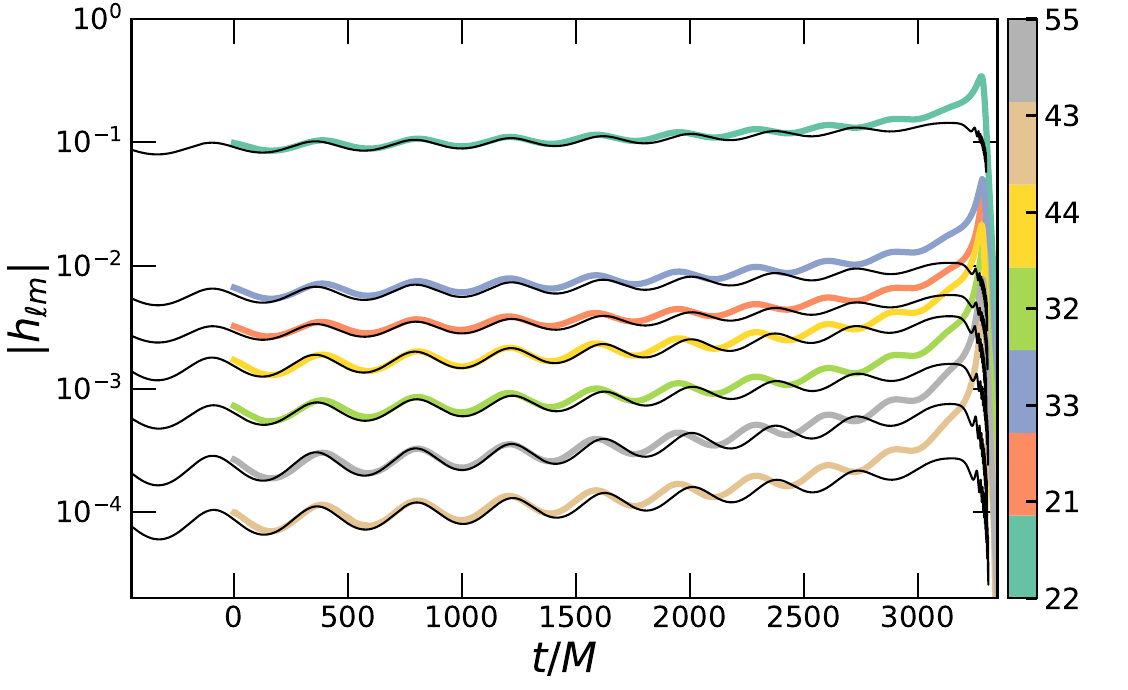}
    \includegraphics[width=0.462\textwidth]{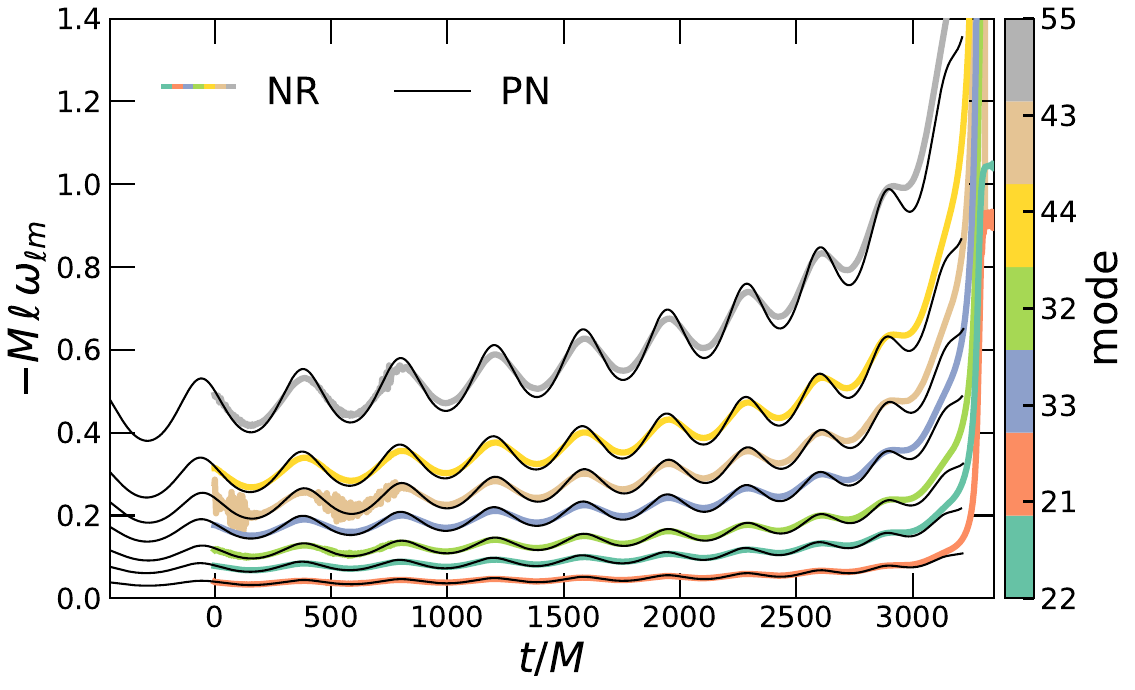}
    \caption{Amplitude and frequency of selected modes from an eccentric NR simulation (SXS:BBH:1364) together with an eccentric PN model are plotted. The eccentricity (mean anomaly) measured roughly $\sim7$ orbits before the merger at a reference frequency of $\sim 0.075$ is 0.044 (2.144) and the binary's mass ratio ($q$) is $2$ (see Table I of \cite{hinder-2018} for details). The PN model is evolved assuming an initial eccentricity of $e_0=0.108$, and mean anomaly of $l_0=-2.495$ (measured at an initial frequency of $x_0=0.045$ using the prescription of \cite{Tanay:2016zog}). Since the mode amplitudes and frequencies are in reasonable agreement for PN and NR waveforms in a time window of ($1000M$, $2000M$) it should be possible to perform hybridization in this window.}
    \label{fig:pn_nr_comparison}
\end{figure*}

The target waveforms used in these analyses have been computed by matching the post-Newtonian inspiral waveforms for individual modes \cite{Boetzel:2019nfw, Ebersold:2019kdc, Tanay:2016zog, Moore:2016qxz} with those extracted from eccentric numerical relativity simulations of the SXS Collaboration \cite{hinder-2018} following extensive comparisons involving waveforms due to the two approaches performed here (see Fig.~\ref{fig:pn_nr_comparison}). Target waveforms (or hybrids as we refer them through the paper) being longer in length compared to NR simulations prove to be critical in accessing the impact of eccentricity and eccentricity-induced corrections to the higher modes in the entire mass range accessible to ground-based detectors such as LIGO and Virgo. A set of 20 eccentric hybrids have been constructed with varying initial eccentricity in the range $0.1\lesssim e_0\lesssim 0.4$ and with mass ratios $q$ = $1, 2, 3$. All the hybrids have a start frequency of $x_0$ = $0.045$ and typically have 30-40 orbital cycles before the merger (see Table~\ref{tab:nrsims} for the details).\\

Finally, a fully analytical dominant mode ($\ell, |m|$) = (2, 2) model obtained by matching an eccentric post-Newtonian (PN) inspiral with a quasi-circular prescription for the merger-ringdown phase calibrated against the 20 eccentric hybrids constructed here is presented. The method used for construction of the model and other relevant details are discussed in Sec. \ref{sec:22-mode-model}. The performance of the model against hybrids can be assessed from the plots presented in Fig.~\ref{fig:td-model-q123-amp} as well as from the mismatch plots displayed in the top-right panel of Fig.~\ref{fig:match_ecc_temp}. Since all 20 hybrids were utilised in calibrating the model, these models are tested against an independent family of eccentric waveforms [eccentric, non-spinning, inspiral, Gaussian-process merger approximant (ENIGMA)~\cite{Chen:2020lzc}], which in turn are calibrated to NR simulations. The mismatch plots with ENIGMA are shown in the bottom-right panel of Fig.~\ref{fig:match_ecc_temp}. For comparison, mismatches with quasi-circular \textsc{SEOBNRv4}~\cite{PhysRevD.95.044028} templates are also shown in the top/bottom left panel of Fig.~\ref{fig:match_ecc_temp}. Both the target and template waveforms used in Fig.~\ref{fig:match_ecc_temp} involve only the dominant (2,2) mode.\\ 

While a higher mode model can be constructed following the methods used in constructing the dominant mode model discussed in Sec.~\ref{sec:22-mode-model}, one may simply use the prescription for the (2, 2) mode model to combine an (eccentric) inspiral and a (quasi-circular) merger-ringdown prescription to obtain an \textit{ad hoc} higher mode (HM) model. Such a model is discussed in Appendix~\ref{sec:hm-model} and its performance against hybrids including higher modes is displayed in Fig.~\ref{fig:hm_model} for three mildly inclined systems (10$^{\circ}$, 20$^{\circ}$, 30$^{\circ}$). This HM model, in addition to the dominant mode, includes all $\ell=m$ modes up to $\ell=5$. 
\\      

This paper is structured as follows: In Sec.~\ref{sec:methods} we discuss the PN and NR inputs as well as the method for combining them for constructing the hybrids. Section~\ref{sec:systematics} discusses the impact of ignoring the presence of orbital eccentricity on detection and parameter estimation of gravitational waves from eccentric BBHs. In particular, Sec.~\ref{sec:match} discusses overlaps (or match maximized over time and phase shifts) between our target hybrids and quasi-circular IMR waveform templates and in Sec.~\ref{sec:pe_systematics}, we discuss systematic biases in parameter estimation studies due to the neglect of eccentricity and eccentricity-induced corrections to the higher modes in recovery templates. Section \ref{sec:22-mode-model} describes the method used for construction of the dominant mode model. Finally, we conclude our work in Sec.~\ref{sec:disc} and summarize our findings. 

\section{Methodology}
\label{sec:methods}

\subsection{PN inspiral and NR inputs}
\label{sec:pn_nr_inputs}
\begin{table}
\begin{tabular}{llllll}
  \hline
  \hline
  Count & Simulation Id & $q$ & $e_0$ & $l_0$ & $N_{\rm orbs}$ \\
\hline
1 & SXS:BBH:1132 & 1 & 0.000 & 2.852 & 53.3\\
2 & HYB:SXS:BBH:1355 & 1 & 0.127 & 2.739 & 40.8\\
3 & HYB:SXS:BBH:1356 & 1 & 0.163 & 1.606 & 40.0\\
4 & HYB:SXS:BBH:1357 & 1 & 0.222 & -1.020 & 36.1\\
5 & HYB:SXS:BBH:1358 & 1 & 0.226 & -2.937 & 35.7\\
6 & HYB:SXS:BBH:1359 & 1 & 0.227 & 1.850 & 36.3\\
7 & HYB:SXS:BBH:1360 & 1 & 0.302 & 0.730 & 31.2\\
8 & HYB:SXS:BBH:1361 & 1 & 0.305 & 1.146 & 31.0\\
9 & HYB:SXS:BBH:1362 & 1 & 0.372 & -0.726 & 25.5\\
10 & HYB:SXS:BBH:1363 & 1 & 0.376 & 0.385 & 25.3\\
11 & HYB:SXS:BBH:1167 & 2 & 0.000 & 1.308 & 48.4\\
12 & HYB:SXS:BBH:1364 & 2 & 0.108 & -2.495 & 46.3\\
13 & HYB:SXS:BBH:1365 & 2 & 0.145 & -1.116 & 44.9\\
14 & HYB:SXS:BBH:1366 & 2 & 0.218 & 0.096 & 39.6\\
15 & HYB:SXS:BBH:1367 & 2 & 0.220 & -0.964 & 40.5\\
16 & HYB:SXS:BBH:1368 & 2 & 0.222 & -1.553 & 40.3\\
17 & HYB:SXS:BBH:1369 & 2 & 0.367 & -2.489 & 28.3\\
18 & HYB:SXS:BBH:1370 & 2 & 0.367 & 0.754 & 28.8\\
19 & HYB:SXS:BBH:1221 & 3 & 0.000 & 2.461 & 56.8\\
20 & HYB:SXS:BBH:1371 & 3 & 0.133 & -1.757 & 52.8\\
21 & HYB:SXS:BBH:1372 & 3 & 0.212 & -2.101 & 48.7\\
22 & HYB:SXS:BBH:1373 & 3 & 0.214 & -2.655 & 48.6\\
23 & HYB:SXS:BBH:1374 & 3 & 0.359 & -2.953 & 35.0\\
\hline



\end{tabular}
\caption{Hybrids constructed by matching NR simulations from the SXS Collaboration and available PN prescriptions for BBHs in eccentric orbits. SXS simulation ID numbers are retained for identification with NR simulation used in the process of construction of the hybrids. Each hybrid starts at an averaged orbital frequency of $x_0=0.045$ where eccentricity ($e_0$) and mean anomaly ($l_0$) are estimated. Mass ratio ($q$) and number of orbits prior to the merger are also listed. $N_{\rm orb}$ has been computed by taking the phase difference between the start of the waveform and the peak of the $(2, 2)$ mode amplitude. Frequency parameter $x = ( \pi M f )^{2/3}$. The NR simulation SXS:BBH:1132 is longer than the hybrids constructed here and hence the NR data are directly used.}
\label{tab:nrsims}
\end{table}

Spin-weighted spherical harmonic modes of the inspiral waveforms from the PN theory, in terms of an amplitude and orbital phase, can be written as follows:
\begin{equation}
\hlm^\mathrm{PN}(t) = {2 G M \eta x\over c^2 D} \sqrt{\frac{16\pi}{5}} \,\hat{H}_{\ell m} \, e^{-i\, m \varphi_\mathrm{orb}(t)},
\label{eq:pn_modes}
\end{equation}
where $\hat{H}_{\ell m}$ is the amplitude of a given ($\ell, m$) mode and $\varphi_\mathrm{orb}(t)$ represents the binary's orbital phase. Symbols $M$ and $D$ represent the binary's total mass and its distance from the observer, while $\eta$ is given by the ratio of the binary's reduced mass to total mass. Unless explicitly mentioned, we work with $G$=$c$=$1$ and set $M$=$1\,M_{\odot}$ and $D$=$1\,{\rm Mpc}$.\\ 
\\
The PN expressions for mode amplitudes ($\hat{H}_{\ell m}$) contributing up to 3PN order for binaries with non-spinning compact components in \emph{quasi-circular} orbits have been computed in Refs.~\cite{BIWW:1996, ABIQ:2004, Kidder:2007, blanchet-2008, kidder-2008}. In fact, some of the leading modes are actually known (or can easily be computed using available inputs) with higher PN accuracy and contribute to relevant modes at higher PN orders \cite{MQ_4pn, MO_3p5pn, CQ_3pn, MQ_3p5pn}. The orbital phase accurate to 3.5PN order has been computed in \cite{BIJ:2001, BFIJ:2002, BDEI:2004} (see Ref.~\cite{Blanchet:LivRev} for a review on the subject). Expressions for mode amplitudes  constituting 3PN inspiral waveforms assuming binaries with non-spinning compact objects in \emph{quasi-elliptical} orbits have been computed in \cite{Boetzel:2019nfw, Ebersold:2019kdc, Mishra:2015bqa} by employing generalised 3PN quasi-Keplerian representation of \cite{damour-1985, memmesheimer-2004}. On the other hand a 3.5 PN prescription for the orbital phase for an eccentric system has been presented in Ref.~\cite{koenigsdoerffer-2006} and is based on phasing formulation of \cite{damour-2004} and generalised quasi-Keplerian representation of \cite{memmesheimer-2004}. While the phasing of \cite{koenigsdoerffer-2006} includes contributions due to the binary's reactive dynamics to relative 1PN order (or absolute order of 3.5PN of phase), Ref.~\cite{Moore:2016qxz} extends these results to the relative 3PN order under the assumption that the binary's initial orbital eccentricity is small ($e_0\lesssim0.2$). In another effort \cite{Tanay:2016zog} the results of Ref. \cite{yunes-2009} were extended to 2PN order and included eccentricity to $\bigO(e^6)$. While the results of Ref. \cite{Moore:2016qxz} should capture relativistic dynamics better (being more accurate in the PN sense), the results of Ref. \cite{Tanay:2016zog} should be applicable to systems with larger eccentricities. \\  
\\

The NR simulations (describing the non-perturbative, late-time evolution of eccentric BBH mergers) used here have been performed using the Spectral Einstein Code \cite{Ossokine:2013zga, Boyle:2019kee}
developed by the SXS 
Collaboration and are publicly available~\cite{SXS:catalog}. A set of 20 eccentric simulations with varying initial eccentricities ($e\leq0.2$) and mass ratios ($q=1, 2, 3$) were first presented in \cite{hinder-2018}.\footnote{See Table I of \cite{hinder-2018} for other relevant properties of these simulations.}

Since the NR simulations we intend to compare PN results with include waveforms with eccentricities as large as 0.2, we have made use of the results of \cite{Tanay:2016zog} while comparing the phase (and the angular frequency), due to its ability to probe larger eccentricities compared to the ones presented in Ref. \cite{Moore:2016qxz}. As far as the amplitude comparisons are concerned we make use of the PN mode amplitudes computed in \cite{Ebersold:2019kdc, Boetzel:2019nfw}. We present the results of these comparisons in the next section.

\begin{figure*}[htbp!]
    \includegraphics[width=0.9\textwidth]{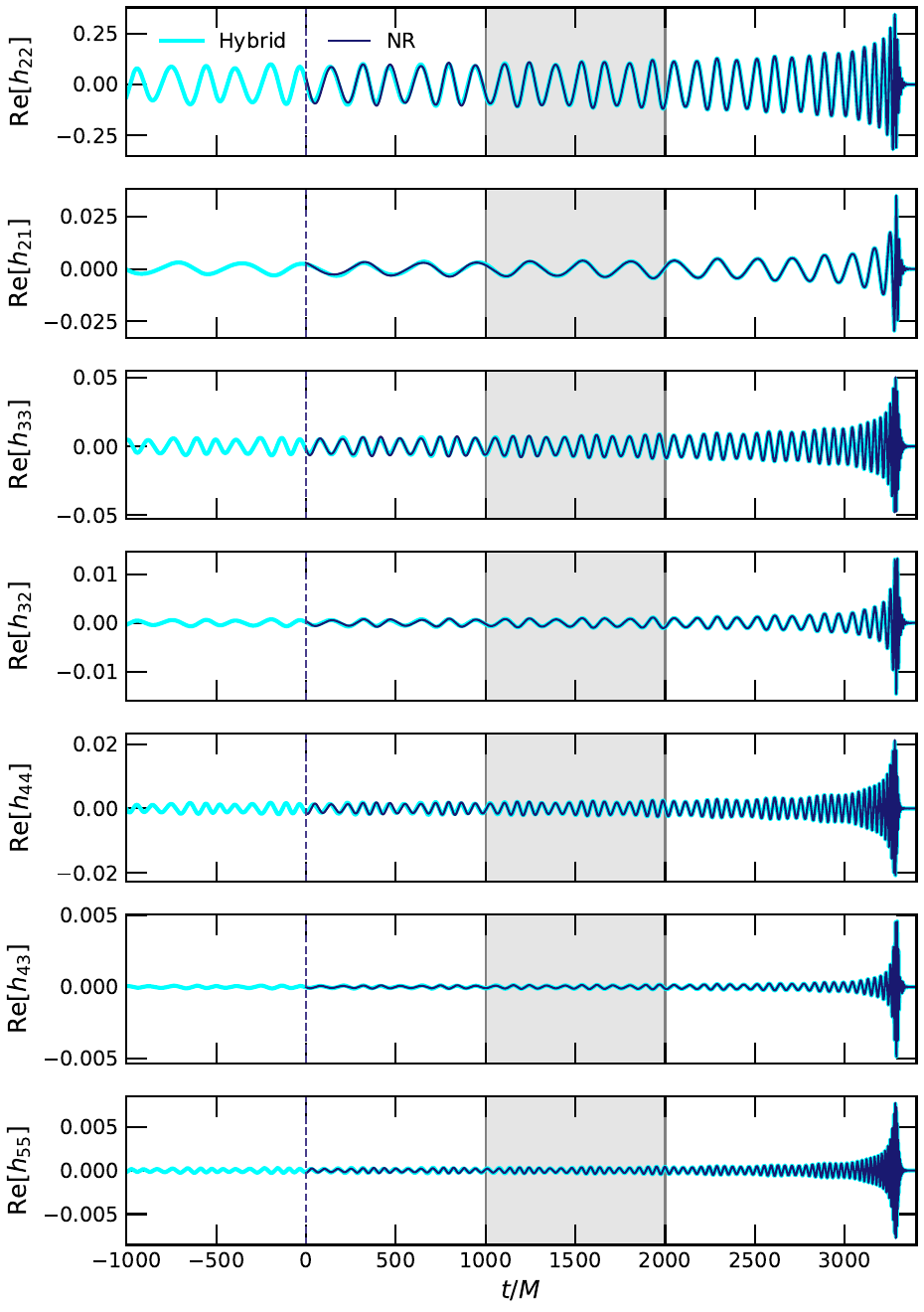}
    \caption{PN-NR hybrid waveform corresponding to NR simulation SXS:BBH:1364, an asymmetric mass binary with mass ratio $q$=$2$. The initial eccentricity of the constructed waveform is $e_{0}=0.108$ at $x_{\rm low}=0.045$. The blue dotted line marks the beginning of the NR waveform and the shaded grey region $t \in (1000M, 2000M)$ shows the matching window where hybridization was performed. Overlapping hybrid and NR waveforms on the left of the matching window hint at the quality of hybridization performed here.}
    \label{fig:hybridplot}
\end{figure*}
\subsection{PN-NR comparison}
\label{sec:pn_nr_comparison}

Figure~\ref{fig:pn_nr_comparison} compares the waveform data of a NR simulation (SXS:BBH:1364) and related results from the PN theory for a number of relevant modes. The colored lines represent the NR data, mimicked closely in first few GW cycles by the PN results that are displayed as black lines. Close agreements between the PN and NR waveforms in the inspiral part of the signal allow for hybridization, as discussed in the next section. However, we choose to ignore the modes whose amplitudes relative to the $\ell$=$2$, $|m|$=2, or simply 22 mode are smaller by a factor of $10^{-3}$, as they may not be relevant for our purposes given their small amplitudes. Additionally, we also demand that data for each mode in the hybridization window should be relatively clean. These two conditions limit the number of modes that are to be included in the hybrids to ($\ell, |m|$) =  (2,2), (2,1), (3,3), (3,2), (4,4), (4,3), and (5,5). Note also that the $m=0$ modes (leading to non-linear GW memory; also known as dc modes) are not included, as they would not impact GW detection and parameter estimation analyses \cite{favata-2009, Pollney:2010hs, Ebersold:2019kdc} and are not extracted accurately in NR simulations. On the other hand, the $m\neq0$ mode includes contributions from what is called oscillatory memory, which becomes relevant in the late stages of binary evolution \cite{favata-2009, Ebersold:2019kdc}. These are included at 3PN level in the PN inspiral waveforms of \cite{Ebersold:2019kdc} that are used in constructing the target hybrids.
\subsection{Construction of hybrid waveforms}
\label{sec:hybrid_waveforms}

\begin{figure*}[t!]
\centering

    \includegraphics[width=0.32\textwidth]{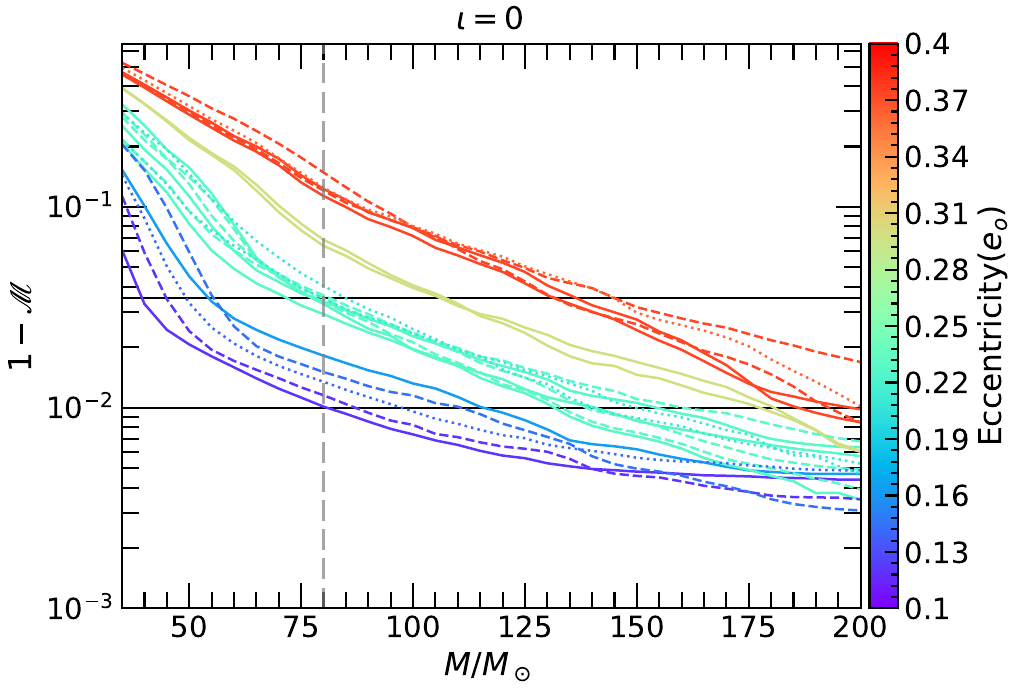}
    \includegraphics[width=0.32\textwidth]{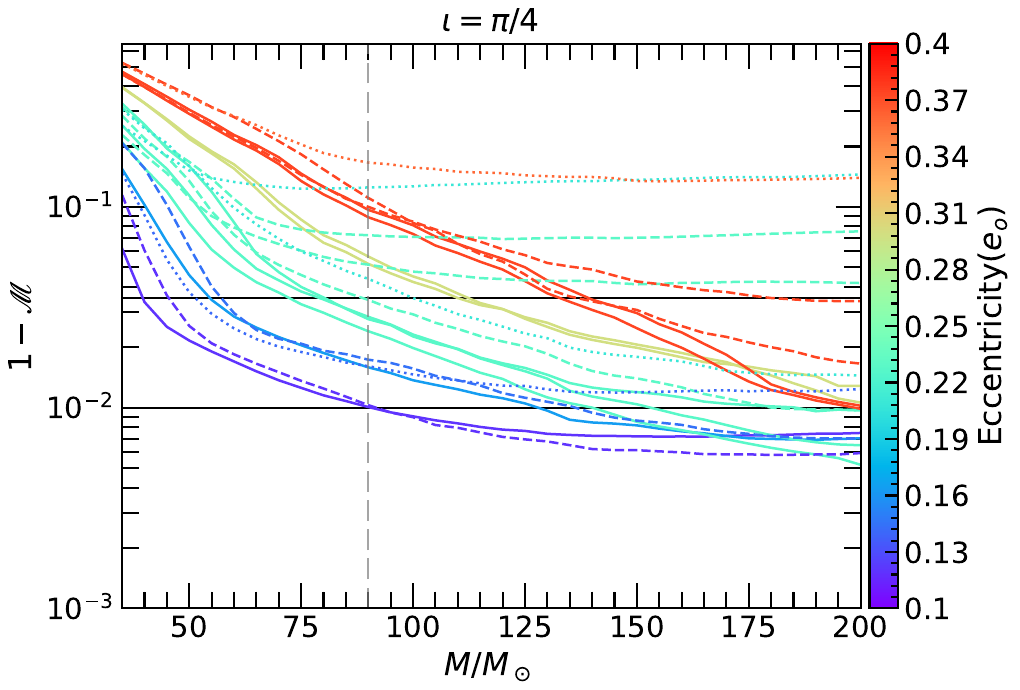}
    \includegraphics[width=0.32\textwidth]{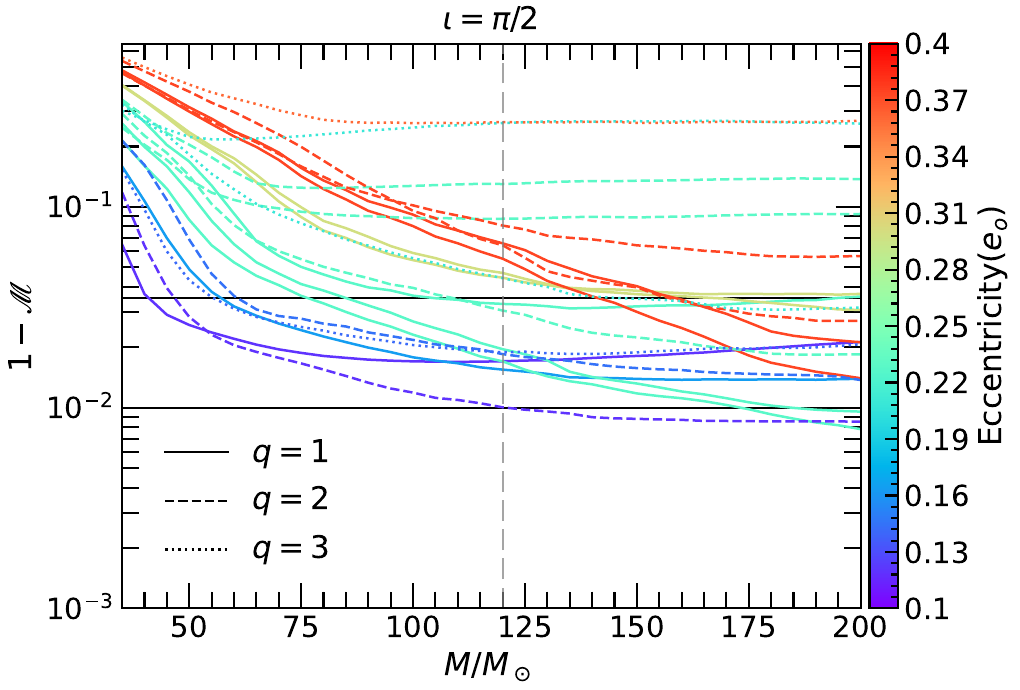}
    \includegraphics[width=0.32\textwidth]{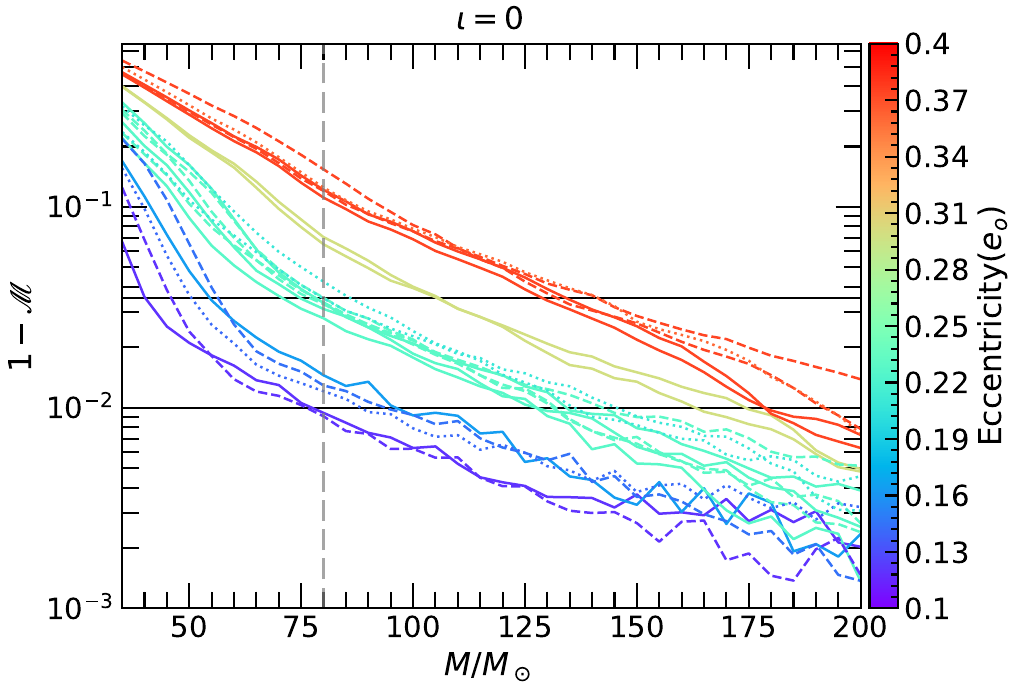}
    \includegraphics[width=0.32\textwidth]{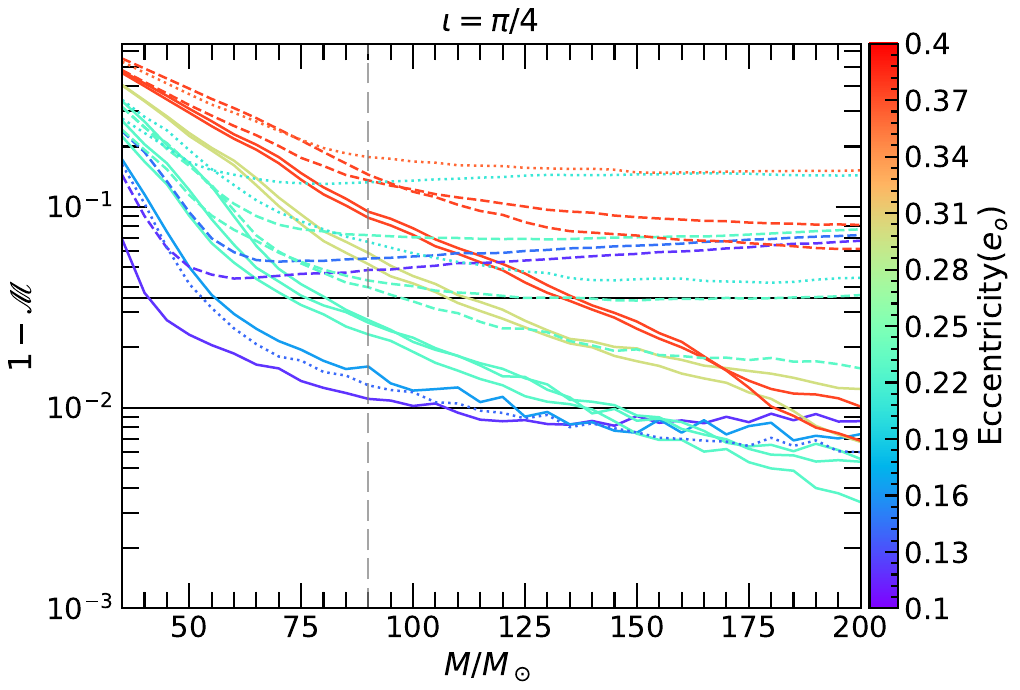}
    \includegraphics[width=0.32\textwidth]{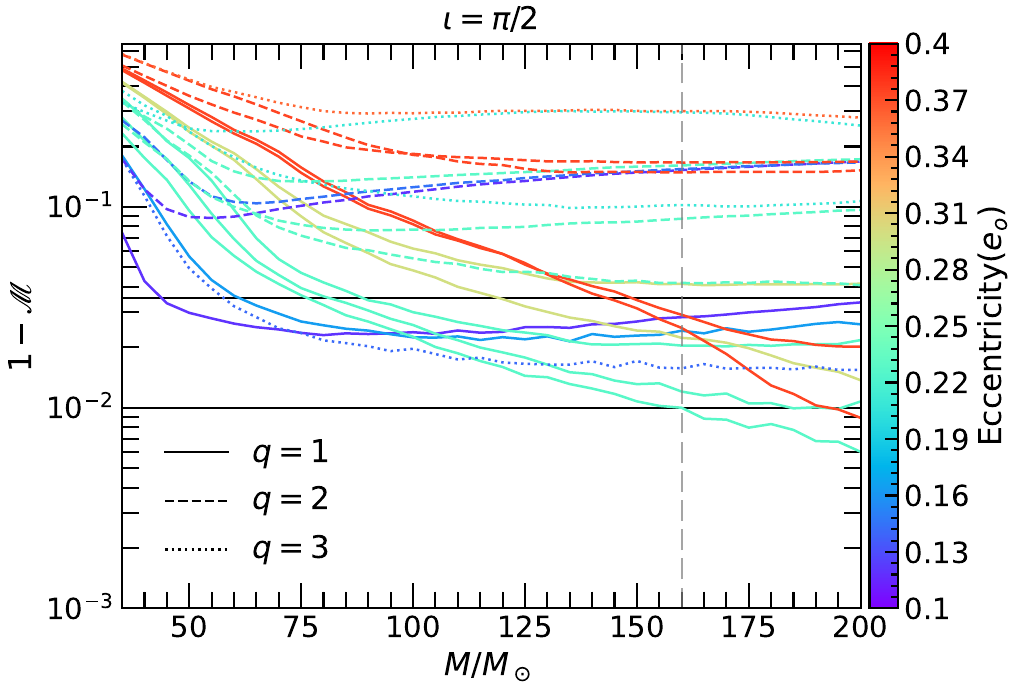}
    \caption{Mismatches between quasi-circular waveforms of \textsc{SEOBNRv4HM}~\cite{Cotesta:2020qhw} (top) and \textsc{IMRPhenomXHM}~\cite{Garcia-Quiros:2020qpx} (bottom) and the eccentric hybrids for three representative orbit inclinations. The two horizontal lines report 96.5\% and 99\% agreement, respectively. The eccentricity values displayed with color bars are computed at $x_0 = 0.045$ for all hybrids. Grey (dashed) vertical lines indicate the total mass below which mismatches become $>1\%$ irrespective of the binary's orbital eccentricity, mass ratio, and orbital inclination.}
    \label{fig:match}
\end{figure*}

 Complete inspiral-merger-ringdown waveforms are constructed by matching PN and NR waveforms for individual modes in a region where the PN prescription closely mimics the NR data following the method of Ref. \cite{Varma:2016dnf}. These are traditionally referred to as ``hybrids" and are used as targets for modeling and data analysis purposes. As discussed in \cite{Varma:2016dnf}, construction of hybrids including higher modes (in the circular case) is possible by performing at least two rotations (and a time shift) so as to align the frames in which PN/NR waveforms are defined.\footnote{It is assumed that the third Euler angle can easily be fixed in the direction of the binary's total angular momentum (see Fig.~2 and the discussions in Sec. III C of ~\cite{Varma:2016dnf}).} We simply extend this argument to the case of eccentric orbits, assuming that the effect of marginalising over parameters such as eccentricity and mean anomaly will not significantly affect the hybridization. The prescription for construction of hybrids is discussed in detail in \cite{Varma:2016dnf} and we reproduce some of the steps here for completeness.\\ 
\\ 
A least-squares minimization of the integrated difference between the GW modes from the PN and NR waveforms in a time interval ($t_{\rm i}, t_{\rm f}$), in which the two approaches give similar results, is performed and can be defined as  
\begin{equation}
\delta = \mathrm{min}_{t_0,\varphi_0, \psi} \int_{t_{\rm i}}^{t_{\rm f}} dt \sum_{\ell,m} \left|\hlm^\mathrm{NR}(t-t_0) e^{i (m \varphi_0 + \psi)}  - \hlm^\mathrm{PN}(t) \, \right|.
\label{eq:delta}
\end{equation}
where the minimization is performed over a time shift ($t_0$) and the two angles ($\varphi_0, \psi$) as discussed above. The hybrid waveforms are then constructed by combining the NR data with the ``best matched'' PN waveform in the following way:
\begin{equation}
\hlm^\mathrm{hyb}(t) \equiv \, \tau(t) \, \hlm^\mathrm{NR}(t-t_0') \ e^{i(m\varphi_0'+\psi')} + (1-\tau(t)) \, \hlm^\mathrm{PN}(t) ,
\label{eq:hyb}
\end{equation}
where ($t_0'$, $\varphi_0'$, $\psi'$) are the values of ($t_0$, $\varphi_0$, $\psi$) that minimize the integral of Eq. \eqref{eq:delta}. In the above, $\tau(t)$ is a weighting function defined by
\begin{eqnarray}
\tau(t) \equiv \left\{ \begin{array}{ll}
0 & \textrm{if $t < t_{\rm i} $}\\
\frac{t-t_{\rm {i}}}{t_{\rm {f}}-t_{\rm {i}}}  & \textrm{if $t_{\rm{i}} \leq t < t_{\rm{f}} $}\\
1 & \textrm{if $t_{\rm{f}} \leq t$.}
\end{array} \right.
\label{eq:tau}
\end{eqnarray}

 The hybrids corresponding to a representative NR simulation (SXS:BBH:1364) for all relevant modes are shown in Fig.~\ref{fig:hybridplot}.
 The blue dotted line marks the beginning of the NR waveform and the shaded grey region $t \in (1000M, 2000M)$ shows the matching window where hybridization was performed. Overlapping hybrid and NR waveforms outside (on the left of) the matching window hint at the quality of hybridization performed here.\\ 
\\
We construct IMR hybrids corresponding to all 20 eccentric NR simulations listed in Ref. \cite{hinder-2018}. 
These are listed in Table~\ref{tab:nrsims} and the SXS simulation IDs have been retained so as to be able to identify the hybrids with the corresponding NR simulation. All the hybrids have a starting frequency of $x_0$=$0.045$, which is the frequency parameter calculated as $x=$ $( \pi M f )^{2/3}$, and typically have 30-40 orbital cycles before the merger. On the other hand, NR simulations used here evolve over 10-15 cycles before the merger. Being longer in length, these hybrids should prove to be critical in assessing the impact of higher modes and eccentricity in the entire mass range accessible to the ground-based detectors such as LIGO and Virgo. We use these hybrids to demonstrate the impact of eccentricity, as well as of higher modes in the section that follows.
It may be useful to note that, for a system of total mass $M\simeq31M_\odot$, the choice of $x$=$0.045$ corresponds to a frequency of $f\simeq20$Hz. Since we work with a low frequency cut-off of 20 Hz (advanced LIGO design \cite{TheLIGOScientific:2014jea}) for all investigations presented here, with current hybrids we can hope to explore systems heavier than $\sim31M_\odot$; in fact, this motivates the choice of a lower mass limit of $35\,M_{\odot}$ in all our analyses. Note, however, since one may generate inspiral waveforms to arbitrary low frequency, hybrids with lower $x_0$ can be generated with little computational cost.  

\section{Impact of eccentricity and higher modes on detection and parameter estimation}
Here we discuss the impact of ignoring the presence of orbital eccentricity on detection and parameter estimation of gravitational waves from eccentric BBHs. Section \ref{sec:match} presents the computation of the match between the hybrids constructed here and state-of-the-art quasi-circular models \cite{Cotesta:2020qhw, Garcia-Quiros:2020qpx} including the effect of higher modes. Section \ref{sec:pe_systematics}, on the other hand, discusses waveform systematics that may be induced due to neglect of eccentricity in recovery waveforms in a parameter estimation study. The discussions of this section are supplemented by those presented in Appendix~\ref{sec:pe-details}.

\label{sec:systematics}
\subsection{Match}
\label{sec:match}

\begin{figure*}[t]
      \centering
      \includegraphics[trim=30 0 30 10, clip, width=0.32\linewidth]{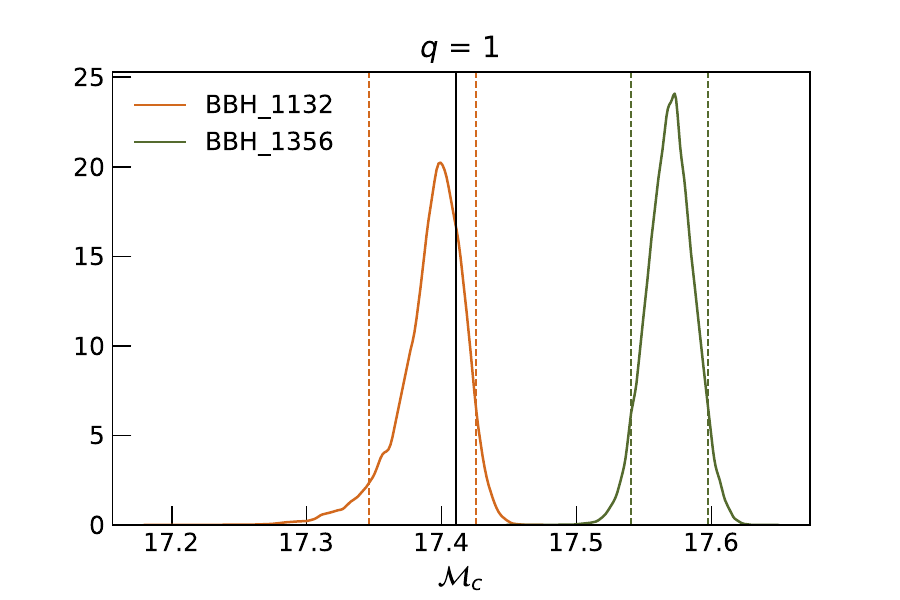}
      \includegraphics[trim=30 0 30 10, clip, width=0.32\linewidth]{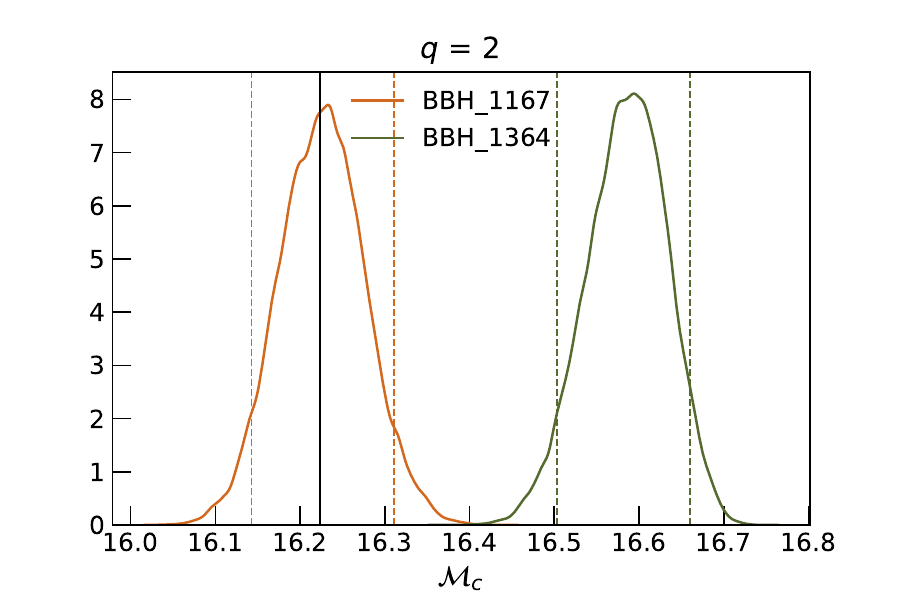}
      \includegraphics[trim=30 0 30 10, clip, width=0.32\linewidth]{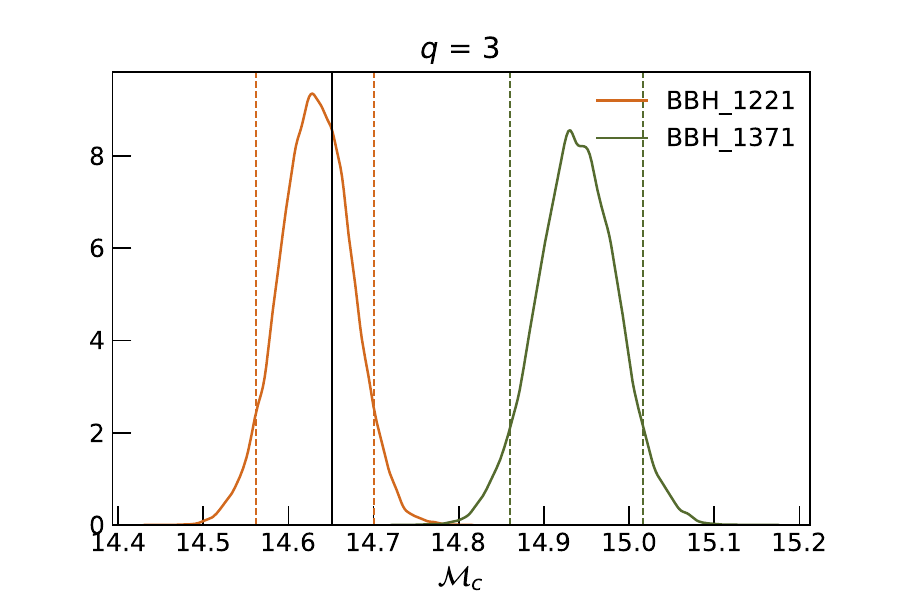}
      \caption{Chirp-mass recovery of circular and eccentric injections using quasi-circular waveforms is shown. Thick black lines denote the injected value of the chirp-mass parameter and the dashed lines denote 90\% credible intervals. The orange (olive) posteriors denote measurement of the chirp mass for the injected circular (eccentric) signal. Both circular and eccentric injections correspond to a BBH of total mass 40$M_{\odot}$, and the eccentric simulations have an orbital eccentricity of $e_{0}\sim0.1$ at 20 Hz. Non-recovery of the injected value of the chirp mass for the eccentric case can be interpreted as the bias induced due to the neglect of eccentricity and associated higher modes.
      }
    \label{fig:pe_systematics}
  \end{figure*}

Figure~\ref{fig:match} displays the mismatch between a set of hybrids including the effect of eccentricity and the higher modes constructed here, and the state-of-the-art quasi-circular waveform families including higher modes \textsc{SEOBNRv4HM}~\cite{Cotesta:2020qhw} (top) and \textsc{IMRPhenomXHM}~\cite{Garcia-Quiros:2020qpx} (bottom). 
As discussed earlier our target hybrids include ($\ell$, $|m|$)=(2, 2), (3, 3), (4, 4), (5, 5), (2, 1), (3, 2), and (4, 3) modes. The \textsc{SEOBNRv4HM}~\cite{Cotesta:2020qhw} templates (quasi-circular)  include ($\ell$, $|m|$)=(2, 2), (3, 3), (4, 4), (5, 5), and (2, 1) modes whereas those of \textsc{IMRPhenomXHM}~\cite{Garcia-Quiros:2020qpx}  include ($\ell$, $|m|$)=(2, 2), (3, 3), (4, 4), (2, 1), and (3, 2) modes.
This means target waveforms may have an additional mode or two depending upon which template is used for recovery. The mismatch plots indicate the need for including eccentricity in gravitational waveforms for analysing BBH systems with masses below 80$M_{\odot}$, irrespective of eccentricity or mass ratio for face-on systems. The mismatches are larger for larger eccentricities. Furthermore, with increasing inclination angle, inclusion of eccentricity in waveforms seems to become important even for heavier systems. For instance, mismatches are $>1\%$ for systems lighter than 90$M_{\odot}$(160$M_{\odot}$) for inclination angle of $\pi/4$ ($\pi/2$). Additionally, as should be clear from middle and right top/bottom panels of Fig.~\ref{fig:match}, with increasing the orbit's inclination with respect to our line of sight, the combined impact of eccentricity and higher modes becomes more apparent.

\subsection{Parameter estimation systematics}
\label{sec:pe_systematics}

\begin{figure*}[ht!]
    \centering
        \includegraphics[width=0.9\textwidth]{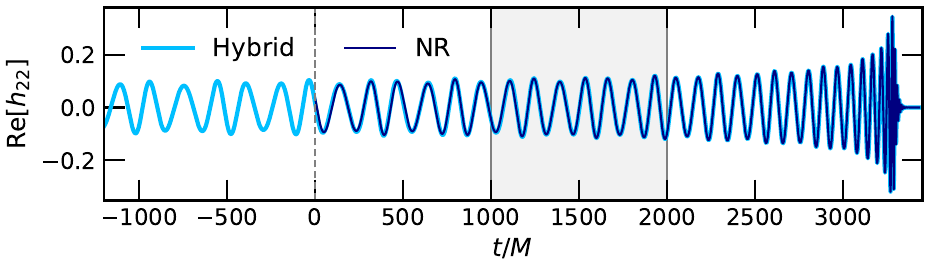}
    \caption{PN-NR hybrid waveform corresponding to NR simulation SXS:BBH:1364, an asymmetric mass binary with mass ratio $q$=$2$, which has been used for constructing the models. The initial eccentricity of the constructed waveform is $e_{0}=0.108$ at $x_{\rm low}=0.045$. The eccentric inspiral waveform used for constructing the target hybrids for modeling are presented in \cite{Tanay:2016zog}.}
    \label{fig:NR-hyb-EccTD}
\end{figure*}

In this section we perform injection studies involving the eccentric, higher mode hybrids constructed above, to assess the biases that are introduced when quasi-circular waveforms are used to recover the injected eccentric signals (hybrids). We use a Bayesian inference method to compute the likelihood and hence construct the posteriors. The posterior probability for a parameter $\Vec{\theta}$ given the data $\Vec{d}$ and a gravitational wave model H, is given by

\begin{equation}
    p(\Vec{\theta}|\Vec{d},H) = \frac{p(\Vec{d}|\Vec{\theta},H) p(\Vec{\theta},H)}{p(\Vec{d}|H)}\,,
\end{equation}
where $p(\Vec{d}|\Vec{\theta},H)$ represents the likelihood, $p(\Vec{\theta},H)$ is the prior, and $p(\Vec{d}|H)$ represents the evidence. Further details regarding the method can be found in Ref. \cite{Biwer:2018osg}. Parameter estimation is performed using the \texttt{PyCBC Inference Toolkit}~\cite{Biwer:2018osg}. We have sampled over the parameter space containing {chirp mass ($\mathcal{M}_c$), symmetric mass ratio ($\eta$), time of coalescence ($t_c$), luminosity distance ($d_L$), phase of coalescence ($\phi_c$), inclination angle ($\iota$), right ascension, and declination}. In our analysis, we have marginalized over the polarization parameter and have put the component spin vectors to zero in the recovery.\\
\\
Figure~\ref{fig:pe_systematics} shows 90\% error bounds in the measurement of the binary's chirp mass ($\mathcal{M}_c$) and are based on a Bayesian analysis performed on simulated data containing injections (hybrids) with distinct mass ratios ($q$=$1, 2, 3$). The total mass of injected signals is assumed to be fixed at 40$M_{\odot}$ and luminosity distance ($d_L$) at 410 Mpc. The value of eccentricity at 20 Hz (starting frequency of the analysis) is $e_0 \sim 0.1$. These are assumed to have non-spinning components and be inclined at $30^{\circ}$ with respect to the binary's orbital angular momentum (also the line of sight). Angular parameters giving the binary's location [$\theta$ and $\phi$ (sky angles)], as well as the polarisation angle ($\psi$), are chosen arbitrarily with values $\pi/6$, $\pi/4$, and $\pi/3$, respectively. Unlike the match computations of  Sec.~\ref{sec:match}, injections here include ($\ell$, $|m|$)=(2, 2), (3, 3), (4, 4), (2, 1), and (3, 2) modes. These are then recovered using circular higher mode waveform \textsc{IMRPhenomXHM} \cite{Garcia-Quiros:2020qpx}, which contains all the above mentioned modes. Additional $(4, 3)$ and $(5, 5)$ modes present in the hybrids have been dropped to have the same set of modes in injection and the recovery waveforms to avoid misinterpretation of the results. The aim here is to observe if the bias in the parameter estimates is purely due to the combined impact of eccentricity and associated higher order modes.
The injected parameter values are indicated by vertical black lines in the figure, while the recovery is shown by posteriors with 90\% error bounds (vertical dashed lines). An injection is assumed to be recovered if the injected value falls within the 90\% bounds. The posteriors in orange denote injections with circular simulations (SXS:BBH:1132, HYB:BBH:1167, HYB:BBH:1221) while those in olive green denote injections with eccentric simulations (HYB:BBH:1356, HYB:BBH:1364, HYB:BBH:1371). \\
\\                                  
It is clear from the figure that eccentric injections are not recovered with the quasi-circular waveform. This indicates that the presence of residual eccentricity of the order $e_0\sim0.1$ and eccentricity-induced corrections to the higher modes in systems entering the ground-based detectors will lead to significant biases in recovering source parameters. This observation should motivate including the effect of orbital eccentricity in dominant and other higher modes in waveforms from compact binary mergers. Recovery of other relevant parameters by the means of corner plots for all three mass ratio cases is displayed in Figs.~\ref{fig:corner_q_1}-\ref{fig:corner_q_3} of Appendix~\ref{sec:pe-details}. This helps us understand correlations between different parameters.

\section{A dominant mode model}
\label{sec:22-mode-model}

\begin{figure*}[ht!]
    \includegraphics[height=4cm, width=0.48\textwidth,keepaspectratio ]{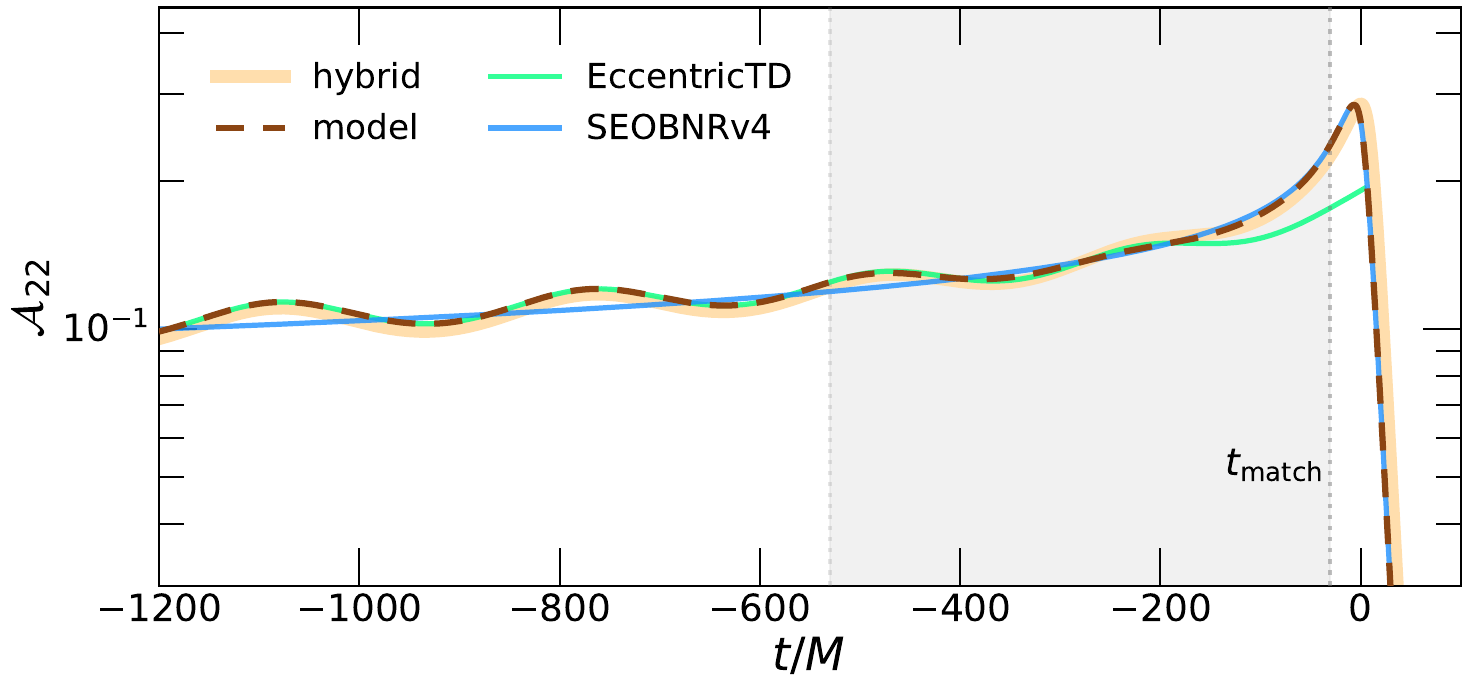}
    \includegraphics[height=4cm,width=0.48\textwidth, keepaspectratio]{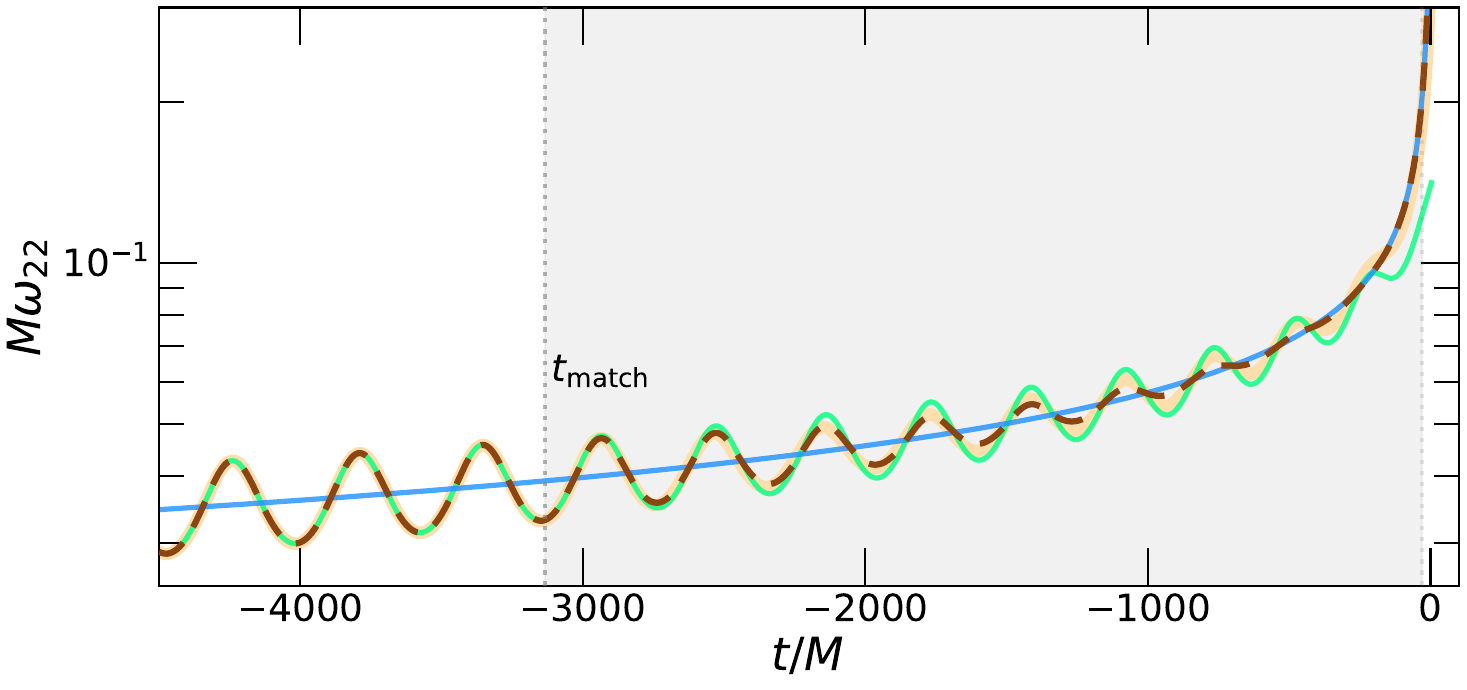}
    \caption{Amplitude and frequency model produced by combining an eccentric inspiral waveform with a circular merger-ringdown waveform. Left: the amplitude transitions smoothly from the inspiral to the merger-ringdown stage inside the shaded region, ending at $t_{\rm match}$. Right: the frequency transitions smoothly from the inspiral to the merger-ringdown stage, starting at $t_{\rm match}$ and ending at $30M$ before merger.}
    \label{fig:td-model-amp}
\end{figure*}

Here we develop a fully analytical dominant ($\ell, |m|$) = (2, 2) mode model obtained by matching an eccentric PN inspiral with a quasi-circular prescription for the merger-ringdown phase; the model is calibrated against a set of 20 eccentric hybrids constructed here. The method used for construction of the hybrids is discussed in Sec.~\ref{sec:methods}. We use another set of hybrids here (constructed by simply matching waveforms of \cite{Tanay:2016zog} with NR simulations of \cite{hinder-2018}) as only a dominant mode target is required. A graphical representation of this hybrid corresponding to the simulation SXS:BBH:1364 is shown in Fig.~\ref{fig:NR-hyb-EccTD}. Overlapping hybrid and NR waveforms outside (on the left of) the matching window hint at the quality of hybridization performed here. The inspiral part of the model is the waveform presented in \cite{Tanay:2016zog}, while the quasi-circular merger-ringdown part is described by the waveform from EOB family discussed in \cite{Cotesta:2020qhw}. Next we discuss the construction of the model and present the analytical prescription for the same.

\subsection{Time-shift}
As described in Sec.~\ref{sec:hybrid_waveforms}, the process of hybridization involves minimization over a time shift. So when producing the amplitude model, we first perform a time shift of the inspiral waveform relative to the circular IMR waveform, because we do not know the exact time to merger. This is done by first setting the merger time for the circular IMR waveform to zero and then time sliding the eccentric inspiral about the merger. We obtain a numerical estimate of the time shift for each target hybrid and denote it by $t_{\rm{shift}}$. Once the time shift is performed, the amplitude and frequency model is generated using the prescription as discussed in Secs. \ref{sec:amplitude model} and \ref{sec:frequency model}, respectively.

\subsection{Amplitude model}
\label{sec:amplitude model}

\begin{figure*}[htbp!]
   \centering
    \includegraphics[width=0.32\textwidth]{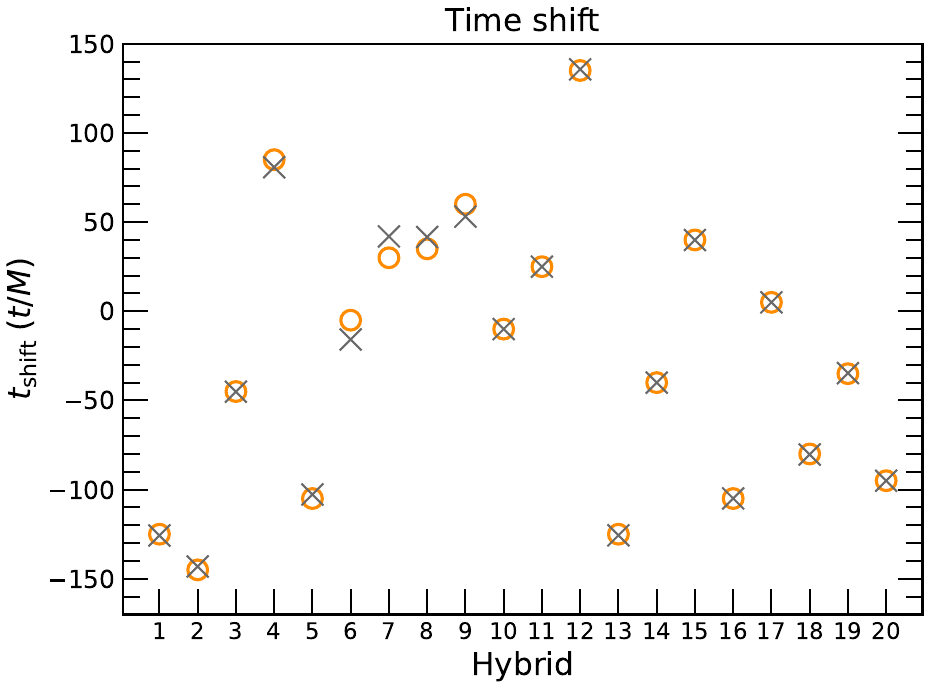}
    \includegraphics[width=0.32\textwidth]{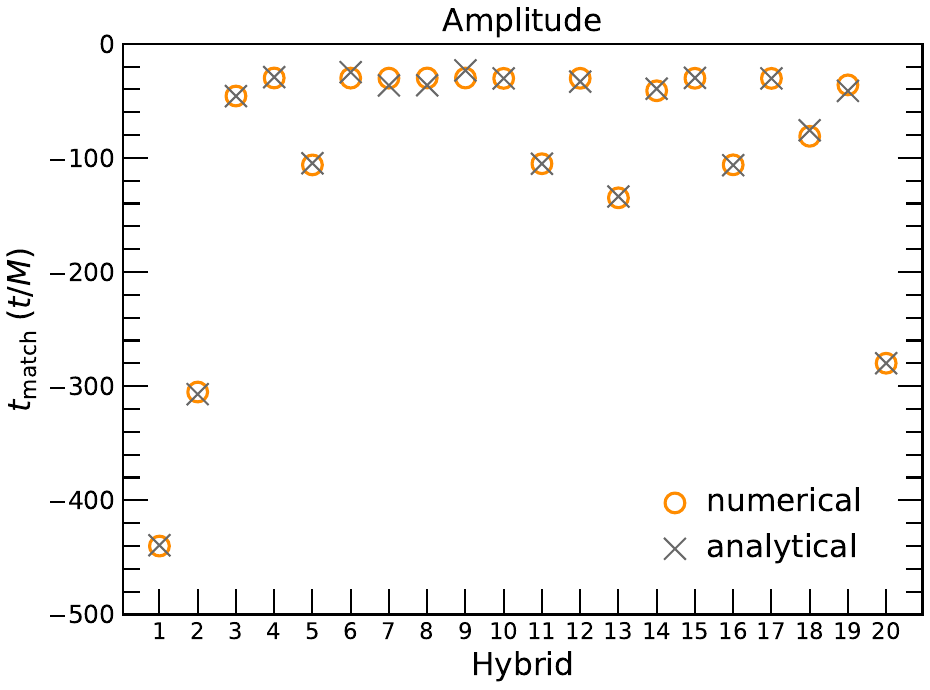}
    \includegraphics[width=0.32\textwidth]{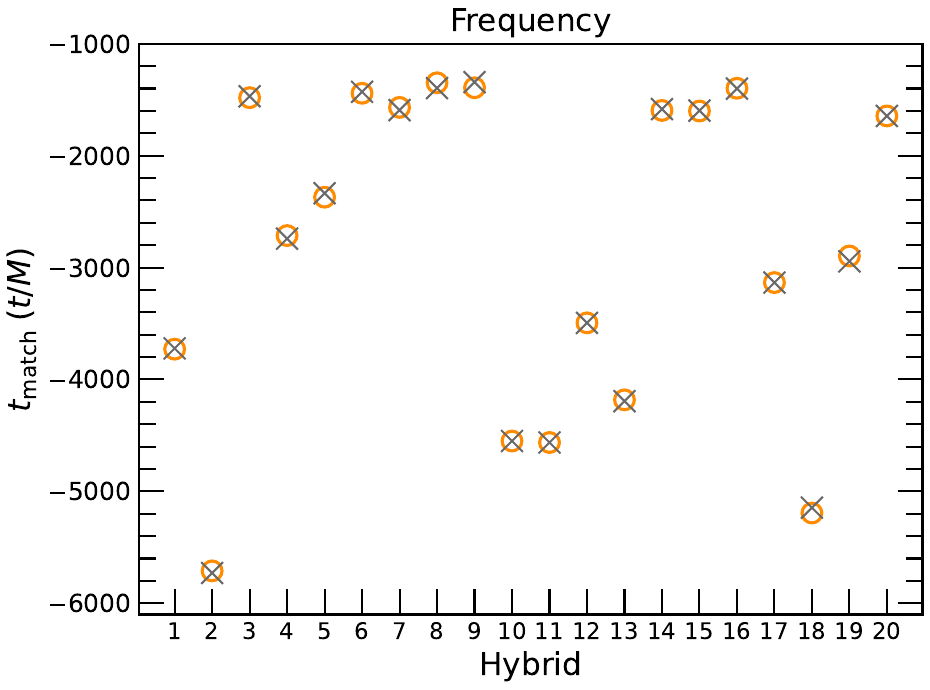}
   \caption{Numerical fits are mapped into the physical parameter space for eccentric systems characterised by the binary's eccentricity, mean anomaly at a reference frequency and the mass ratio parameter $q$ or $\eta$ depending upon the model. Circles represent the numerical data points while crosses represent the value returned by the best-fit model.}
    \label{fig:amplitude_fit}
\end{figure*}

As shown in the plots in Fig.~\ref{fig:pn_nr_comparison}, the waveforms tend to circularize near merger.\footnote{See also the discussion around Fig. 3 of \cite{hinder-2018} which clearly shows all NR simulations become circular $30M$ before the merger.} Hence, in order to model this effect we can suitably join the eccentric inspiral to the circular IMR at an appropriate time $t_{\rm{match}}$. The amplitude model is obtained by joining the eccentric inspiral with the circular IMR using a transition function over a fixed time interval of $500M$ which ends at $t_{\rm{match}}$. Given a target hybrid, we start with a trial choice of $t_{\rm{match}} $ roughly $500 M_\odot$ before the merger and produce the amplitude model as given below,

\begin{equation}
\mathcal{A}_{22}^\mathrm{model}(t) \equiv \, \tau_{\rm{a}}(t) \, \mathcal{A}_{22}^\mathrm{IMR}(t) \  + (1-\tau_{\rm{a}}(t)) \, \mathcal{A}_{22}^\mathrm{inspiral}(t) ,
\label{eq:amp_model}
\end{equation}
where $\tau_{\rm{a}}(t)$ is defined as 

\begin{eqnarray}
\tau_{\rm{a}}(t) \equiv \left\{ \begin{array}{ll}
0 & \textrm{if $t < t_{\rm{i}} $}\\
\frac{t-t_{\rm{i}}}{t_{\rm{f}}-t_{\rm{i}}}  & \textrm{if $t_{\rm{i}} \leq t < t_{\rm{f}} $}\\
1 & \textrm{if $t_{\rm{f}} \leq t$.}
\end{array} \right.
\label{eq:tau_amp}
\end{eqnarray}
We set $t_{\rm{i}} = t_{\rm{match}} - 500M$ and $t_{\rm{f}} = t_{\rm{match}}$ as the bounds of the time interval over which the two waveforms are joined. Figure~\ref{fig:td-model-amp} demonstrates the process. The grey region is the time interval ending at $t_{\rm{match}}$ where the inspiral and circular IMR is joined.\\
\\
After the amplitude model is obtained for a particular choice of trial  $t_{\rm{match}}$, we combine it with the target hybrid phase to obtain the polarizations and then calculate the match with the target hybrid. We then change the trial choice of $t_{\rm{match}}$ by $5M$, bringing it closer to the merger, and repeat the process of producing the amplitude model, and calculating the match. This variation of $t_{\rm{match}}$ is done until roughly $30M$ before merger. We thus obtain a set of match values for varying $t_{\rm{match}}$ and pick the one that has the highest value of match. The corresponding amplitude $t_{\rm{match}}$ is the numerical estimate for a particular target hybrid. We obtain numerical estimates using the same process for all 20 target hybrids.

\subsection{Frequency model}
\label{sec:frequency model}

\begin{figure*}[htbp!]
    \centering
    \includegraphics[width=0.49\textwidth]{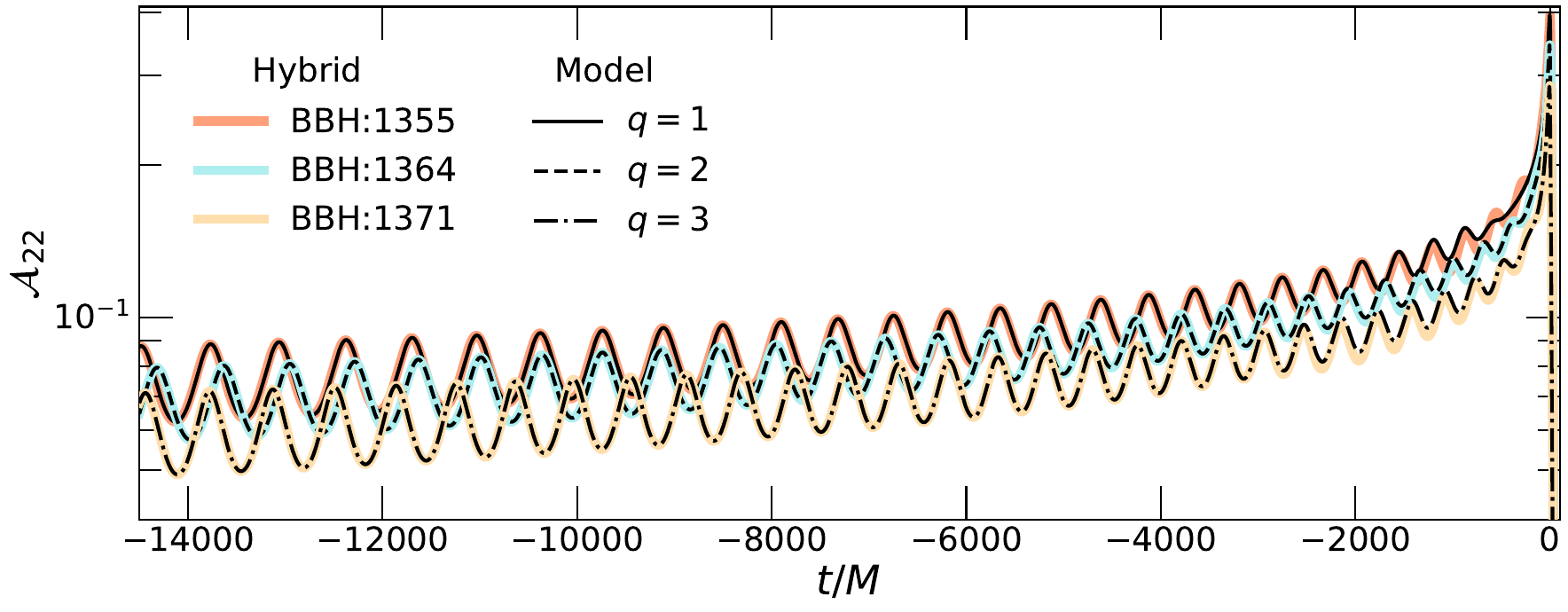}
    \includegraphics[width=0.49\textwidth]{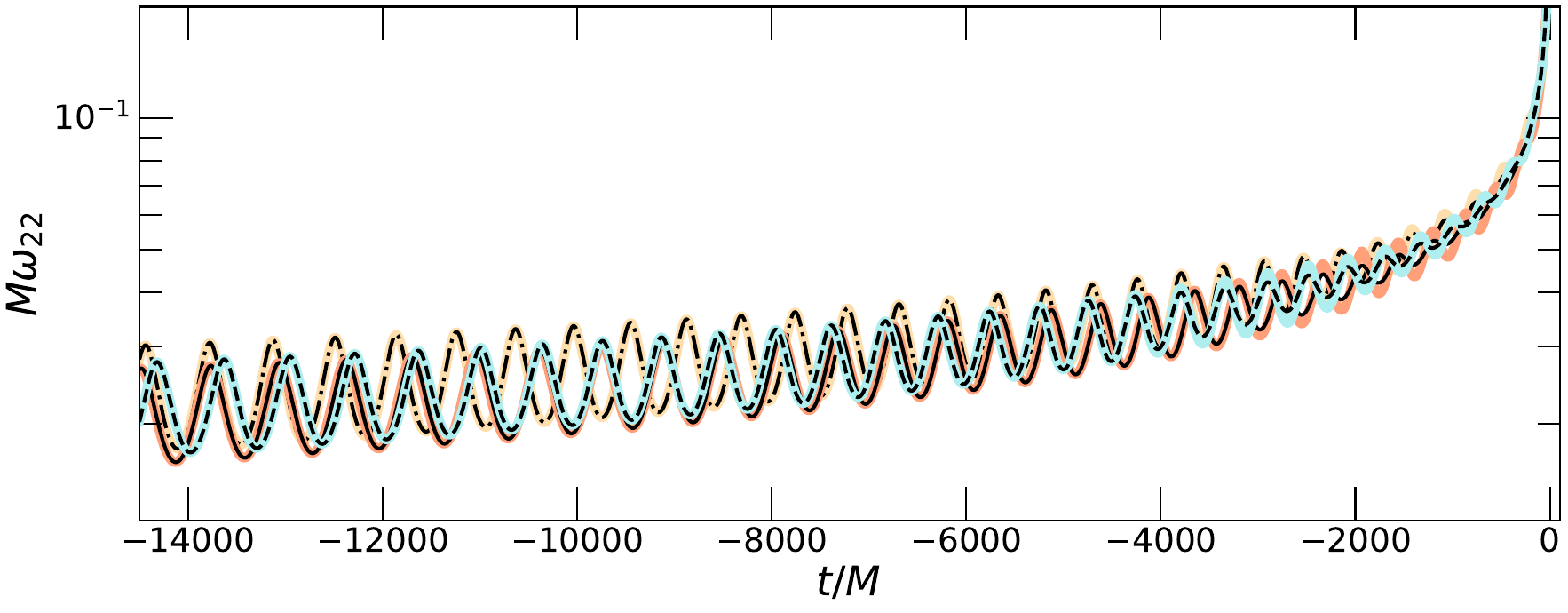}
    \includegraphics[height=4cm, width=0.98\textwidth, keepaspectratio]{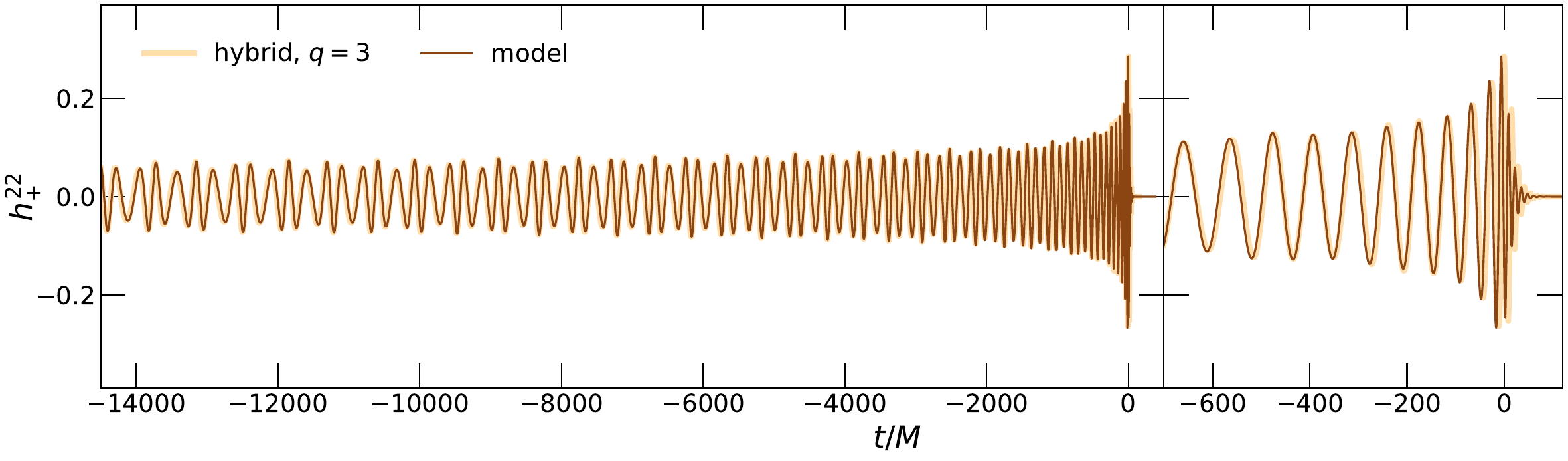}
    \caption{Top: amplitude and frequency of the dominant mode model (constructed out of stitching an inspiral and a merger-ringdown model) plotted together with target hybrids for comparison for three representative low eccentricity simulations. Bottom: one of the polarizations, obtained by combining the amplitude and the frequency model shown in the top panel for the $q=3$ case, is shown as a visual proof of the quality of the model being presented.}
    \label{fig:td-model-q123-amp}
\end{figure*}

For the frequency model, we follow a similar procedure as described in Sec.~\ref{sec:amplitude model} with the only difference being the duration of the time interval where the inspiral frequency is joined with the circular IMR frequency. Once again, similar to the amplitude model procedure, we determine an appropriate $t_{\rm{match}}$ for joining the inspiral frequency with the circular IMR frequency. However, the time interval where the two are joined starts at $t_{\rm{match}}$ and ends at a time close to $30M$\footnote{This choice is motivated by the fact that all NR simulations necessarily circularise $30M$ before the merger\cite{hinder-2018}.} before merger. Just like the amplitude model, we start with the choice of a trial value of $t_{\rm{match}}$ roughly $6000M$ before merger and obtain the frequency model as given below,

\begin{equation}
\omega_{22}^\mathrm{model}(t) \equiv \, \tau_{\rm{a}}(t) \, \omega_{22}^\mathrm{IMR}(t) \  + (1-\tau_{\rm{a}}(t)) \, \omega_{22}^\mathrm{inspiral}(t) ,
\label{eq:frequency_model}
\end{equation}
where $\tau_{\rm{a}}(t)$ is as defined in Eq.~\eqref{eq:tau_amp} with the difference being $t_{\rm{i}} = t_{\rm{match}}$ and $t_{\rm{f}} \lesssim -30M$. Figure~\ref{fig:td-model-amp} demonstrates the process.\\
\\
Once the frequency model is obtained for the choice of trial $t_{\rm{match}}$, we calculate the phase by integrating the frequency model. This is then combined with the amplitude model obtained for the same target hybrid to produce the polarizations and a match with the target hybrid is calculated. We then change the trial choice of $t_{\rm{match}}$ by $1M$, bringing it closer to the merger and repeat the process of producing the frequency model, and calculating the match. Once again, we do this variation until roughly $30M$ before merger to obtain a set of match values for varying $t_{\rm{match}}$ and pick the one that has the highest value of match. The corresponding value of frequency $t_{\rm{match}}$ is the numerical estimate for a particular target hybrid. We obtain numerical estimates for all 20 target hybrids using the same process.

\subsection{Analytical model}
\label{sec:analytical model}

\begin{figure*}[htbp!]
\centering
    \includegraphics[width=0.495\textwidth]{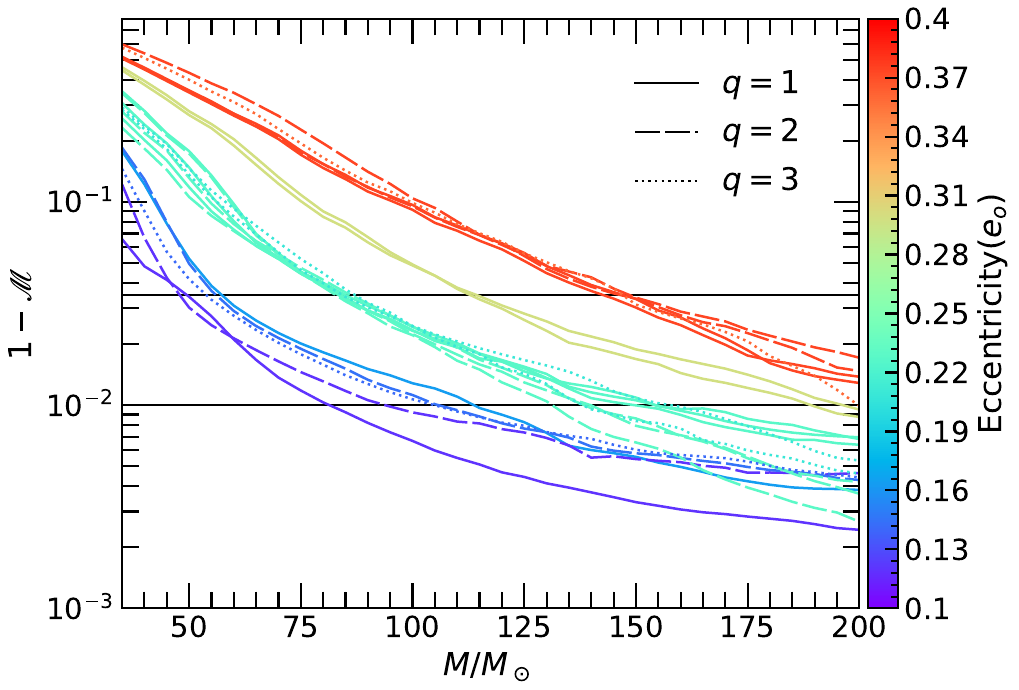}
    \includegraphics[width=0.495\textwidth]{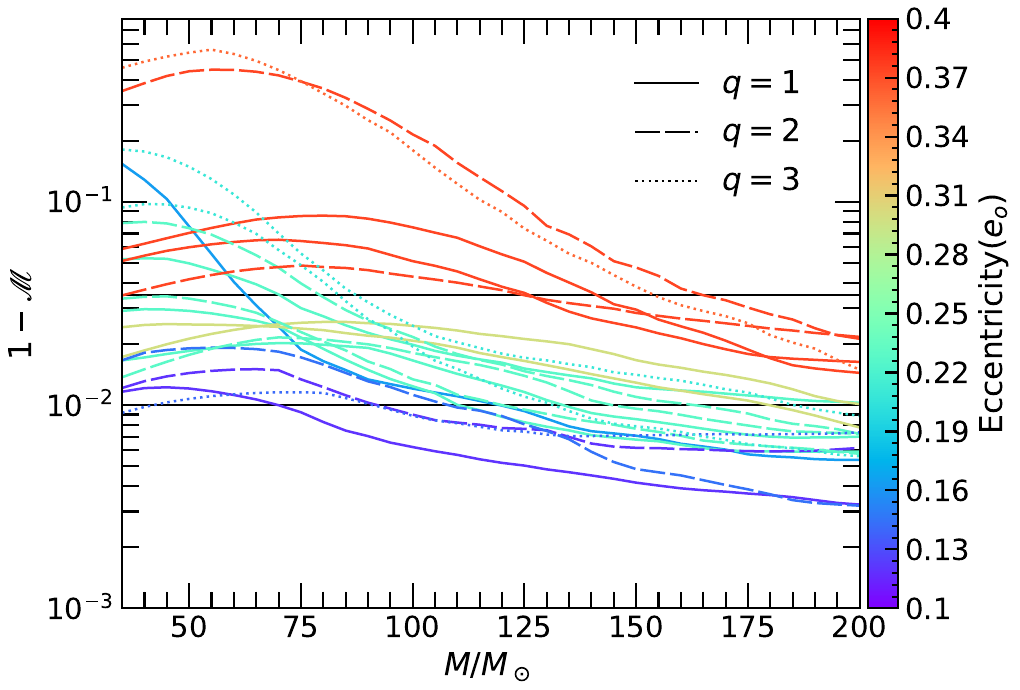}
    
    \includegraphics[width=0.495\textwidth]{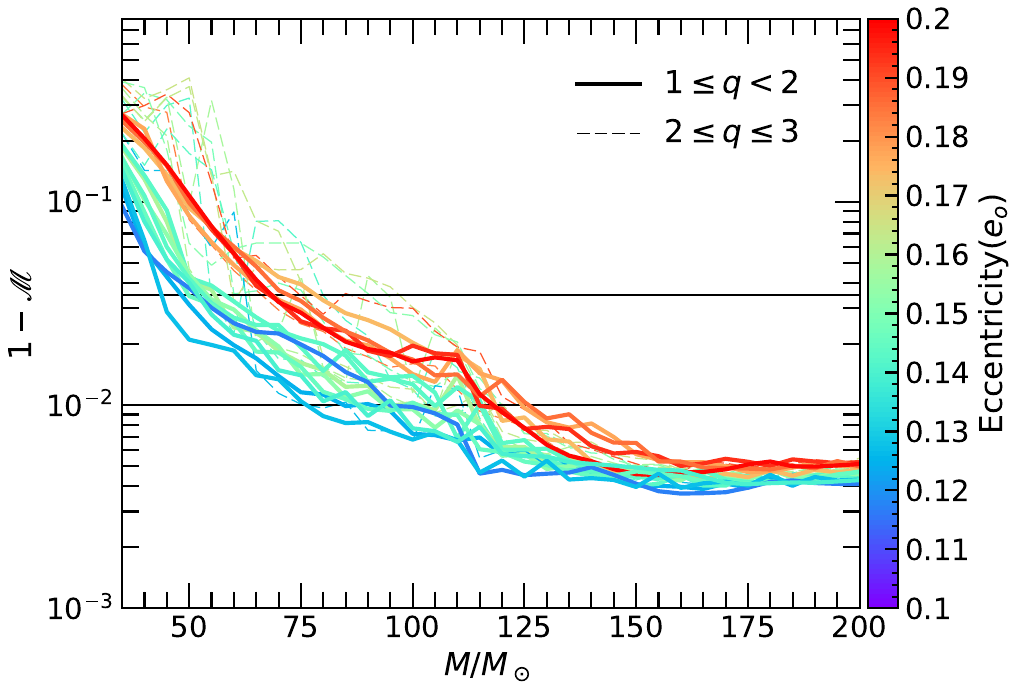}
    \includegraphics[width=0.495\textwidth]{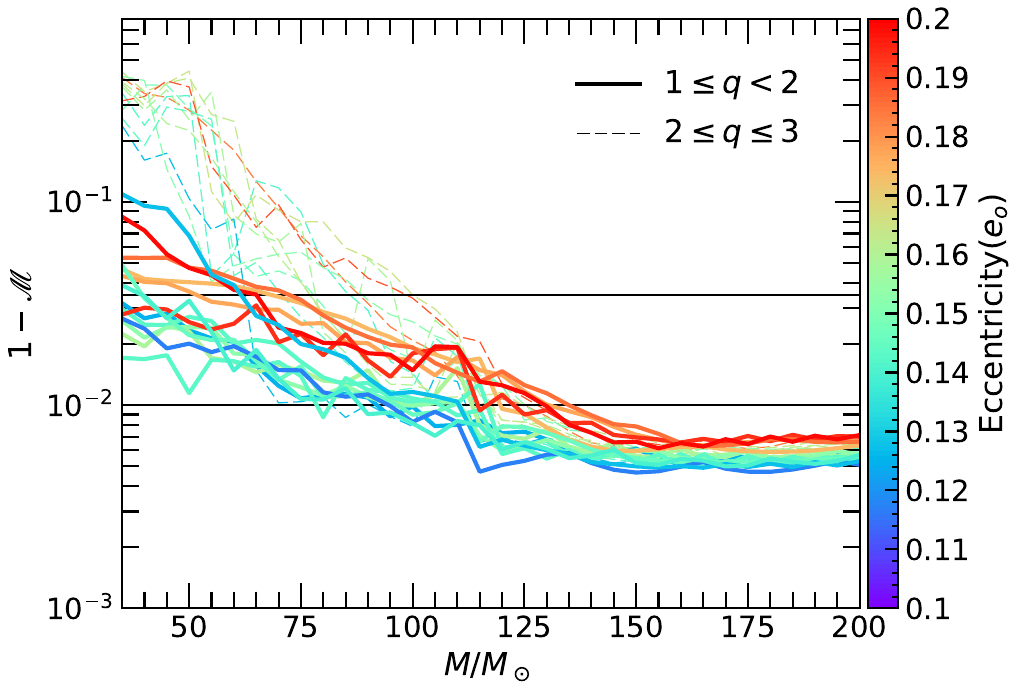}
    \caption{Mismatch of a set of twenty $(2, \pm2)$  mode hybrids constructed here (see Sec.~\ref{sec:22-mode-model}) with a $(2, \pm2)$ mode quasi-circular waveform SEOBNRv4 \cite{PhysRevD.95.044028} (top left) and with the dominant mode model (top right). Mismatch of the dominant mode model with ENIGMA \cite{Chen:2020lzc} (bottom right). For comparison, mismatches of ENIGMA with quasi-circular \textsc{SEOBNRv4}~\cite{PhysRevD.95.044028} templates are also displayed on the bottom left. The two horizontal lines report 96.5\% and 99\% agreement, respectively. The eccentricity values displayed with color bars are computed at $x_0 = 0.045$ for all hybrids.}
    \label{fig:match_ecc_temp}
\end{figure*}

We have described the procedure of producing (numerical) time-domain model fits for the dominant mode model, where we used a set of 20 eccentric hybrids as targets to calibrate our model. For each hybrid, we obtained a numerical estimate for $t_{\rm{shift}}$, amplitude $t_{\rm{match}}$, and frequency $t_{\rm{match}}$. In order to be able to generate waveforms for an arbitrary configuration these numerical fits need to be mapped into the physical parameter space for eccentric systems characterised by binary's eccentricity, mean anomaly at a reference frequency and the mass ratio parameter. In this section, we determine a functional form by performing analytical fits to these numerical estimates. The fitted functions obtained are of the form

\begin{align}
t_{\rm{shift}}\,(q,e,l) &= \sum_{\alpha, \beta, \gamma, \delta} A_{\alpha \beta \gamma \delta} \, e^\alpha \, q^\beta \, \cos( \gamma \, l \,+\, \delta \,e \, l \,+\, a_{\alpha \beta \gamma \delta})
\label{eq:t_shift_analytical}
\end{align}
for time shift, where $A_{\alpha \beta \gamma \delta} = a_{\alpha \beta \gamma \delta} = 0$ for $\alpha + \beta > 4$ and/or $\gamma + \delta > 1$, and $A_{\alpha 0 1 0} = A_{\alpha 0 0 1} = A_{0 \beta 1 0} = A_{0 \beta 0 1} = A_{0 0 \gamma \delta} = a_{\alpha \beta 0 0 } = 0$,

\begin{align}
t_{\rm{match}}\,(\eta,e,l) &= \sum_{\alpha, \beta, \gamma, \delta} B_{\alpha \beta \gamma \delta} \, e^\alpha \, \eta^\beta \, \cos( \gamma \, l \,+\, \delta \,e \, l \,+\, b_{\alpha \beta \gamma \delta})
\label{eq:t_match_amplitude}
\end{align}
for amplitude, where $B_{\alpha \beta \gamma \delta} = b_{\alpha \beta \gamma \delta} = 0$ for $\alpha + \beta > 4$ and/or $\gamma + \delta > 1$, and $B_{\alpha 0 1 0} = B_{\alpha 0 0 1} = B_{0 \beta 1 0} = B_{0 \beta 0 1} = B_{0 0 \gamma \delta} = b_{\alpha \beta 0 0 } = 0$, and

\begin{align}
t_{\rm{match}}\,(\eta,e,l) &= \sum_{\alpha, \beta, \gamma, \delta} C_{\alpha \beta \gamma \delta} \, e^\alpha \, \eta^\beta \, \cos( \gamma \, l \,+\, \delta \,e \, l \,+\, c_{\alpha \beta \gamma \delta})
\label{eq:t_match_frequency}
\end{align}
for frequency, where $C_{\alpha \beta \gamma \delta} = c_{\alpha \beta \gamma \delta} = 0$ for $\alpha + \beta > 5 $ and/or $\gamma + \delta > 1$, and $C_{\alpha 0 1 0} = C_{\alpha 0 0 1} = C_{0 \beta 1 0} = C_{0 \beta 0 1} = C_{0 0 \gamma \delta} = c_{\alpha \beta 0 0 } = 0$. The values for the coefficients $A_{\alpha \beta \gamma \delta}$, $B_{\alpha \beta \gamma \delta}$, $C_{\alpha \beta \gamma \delta}$, $a_{\alpha \beta \gamma \delta}$, $b_{\alpha \beta \gamma \delta}$, and $c_{\alpha \beta \gamma \delta}$ obtained by performing a fit to the numerical values are tabulated in Tables~\ref{tab:shift}-\ref{tab:freq}.\\
\\
Figure~\ref{fig:amplitude_fit} shows comparisons between the numerically obtained values for $t_{\rm{shift}}$, amplitude $t_{\rm{match}}$, and frequency $t_{\rm{match}}$, with the values predicted by our analytical fits. The predictions are within $\pm 12M$ of the numerical estimates for time shift, within $\pm 7M$ for amplitude $t_{\rm{match}}$, and within $\pm 51M$ for frequency $t_{\rm{match}}$. We show amplitude and frequency comparison between the target hybrids and our models for three cases along with the full waveform for the $q$=$3$ case in Fig.~\ref{fig:td-model-q123-amp}.\\

The performance of the model can be assessed from the plots against hybrids presented in Fig.~\ref{fig:td-model-q123-amp}, as well as from the mismatch plots displayed in Fig.~\ref{fig:match_ecc_temp}. Mismatches with our dominant mode model are smaller compared to those with $(2, |2|)$ mode quasi-circular templates of  \textsc{SEOBNRv4}~\cite{PhysRevD.95.044028}, in particular at low mass range for almost all cases. Note that the target hybrids used in these mismatch computations shown in Fig.~\ref{fig:match_ecc_temp} include only the ($2, |2|$) modes so as to assess the actual performance of the dominant mode model. Since all 20 hybrids were utilised in calibrating the model, these models are tested against the independent waveform family (also calibrated against NR simulations) ENIGMA~\cite{Chen:2020lzc}.
For this comparison, we choose to sample a parameter space that is not identical to the calibration set (hybrids). We choose to generate both the target (ENIGMA) and the template (dominant eccentric model) by randomly sampling values of a reference eccentricity ($e_0$), mass ratio ($q$), and reference mean anomaly ($l_0$) in the range $0.1\lesssim e_0\lesssim0.2$, $1\lesssim q\lesssim3$, and $-\pi\leq l_0\leq\pi$, respectively. The mismatch plot obtained is shown in the bottom-right panel of Fig.~\ref{fig:match_ecc_temp}. Additionally, for comparison, mismatches of ENIGMA with quasi-circular \textsc{SEOBNRv4}~\cite{PhysRevD.95.044028} templates are displayed in the bottom-left panel. Clearly, our model seems to do better compared to the circular templates at low mass end ($M\lesssim$100$M_{\odot}$), while mismatches are comparable for heavier systems. The larger mismatches observed at the low mass end might be due to the differences between the inspiral inputs that go in our model and those in ENIGMA~\cite{Chen:2020lzc}. It is also worth noting that these mismatches are significantly larger for mass ratios $q\geq2$ (dashed curves) compared to near equal mass cases (solid curves) irrespective of the choice of eccentricity or mean anomaly. These may be interesting aspects to investigate in a future work or during an independent review of current waveforms.\\ 

Before we conclude this section, we would like to highlight that the effect of higher modes can also be included in a model following the methods used in constructing the dominant mode model presented here. While we defer construction of a higher mode model for a future work, we include a proof of principle demonstration for such constructions in Appendix~\ref{sec:hm-model} where we simply use the prescription for the dominant $(2, 2)$ mode model for combining (eccentric) inspiral and (quasi-circular) merger-ringdown prescriptions for each mode to obtain an \textit{ad hoc} higher mode model. Its performance against hybrids including higher modes is displayed in Fig.~\ref{fig:hm_model} for three mildly inclined systems (10$^{\circ}$, 20$^{\circ}$, 30$^{\circ}$). This HM model, in addition to the dominant mode, includes all $\ell=m$ modes up to $\ell=5$. 

\section{Discussion}
\label{sec:disc}
We started by comparing high accuracy PN inspiral waveforms for compact binaries in eccentric orbits of \cite{Ebersold:2019kdc, boetzel-2017, Tanay:2016zog} with SXS NR data for eccentric BBH mergers presented in \cite{hinder-2018}. Figure~\ref{fig:pn_nr_comparison} compares the waveform data for one particular dataset. Based on this comparison we select a set of modes that are included in the target IMR models (hybrids). The hybrids are constructed for ($\ell$, $|m|$)=(2, 2), (3, 3), (4, 4), (5, 5), (2, 1), (3, 2), and (4, 3) modes. The hybridization procedure is discussed in Sec.~\ref{sec:hybrid_waveforms}. We use these hybrids to show the impact of eccentricity and eccentricity-induced corrections to the higher modes by computing the mismatch with the state-of-the-art quasi-circular waveforms such as \textsc{SEOBNRv4HM} \cite{Cotesta:2020qhw} and \textsc{IMRPhenomXHM} \cite{Garcia-Quiros:2020qpx}. The mismatches are shown in Fig.~\ref{fig:match}. It may be worth noting that target waveforms used in computing mismatches may have an additional mode or two depending upon which template
is used for recovery (see Sec. \ref{sec:match} for details). Subsequently, in Sec.~\ref{sec:pe_systematics}, with an injection analysis we demonstrate it will not be possible to ignore the presence of eccentricity and eccentricity-induced corrections to the higher modes while recovering GW signals using current state-of-the-art quasi-circular waveforms including the effect of higher order modes. Unlike match computations with quasi-circular templates (Fig.~\ref{fig:match}), both the target (hybrids) and the template (\textsc{IMRPhenomXHM} \cite{Garcia-Quiros:2020qpx}) include ($\ell$, $|m|$)=(2, 2), (3, 3), (4, 4), (2, 1), and (3, 2) modes (see a discussion in Sec.~\ref{sec:pe_systematics}).  Finally, in Sec.~\ref{sec:22-mode-model} we develop a fully analytical dominant mode ($\ell, |m|$) = (2, 2) model obtained by matching an eccentric PN inspiral with a quasi-circular prescription for the merger-ringdown phase calibrated against a set of eccentric hybrids. Figure~\ref{fig:td-model-amp} and \ref{fig:amplitude_fit} demonstrate the procedure, while Figs.~\ref{fig:td-model-q123-amp} and \ref{fig:match_ecc_temp} display the performance of the model developed here. Additionally, a simple extension of the dominant mode model to include the effect of higher modes is also presented in Appendix~\ref{sec:hm-model}. This higher mode model, in addition to the dominant mode, includes all $\ell=m$ modes up to $\ell$=$5$.\footnote{In principle, we could also model (2, 1) mode as \textsc{SEOBNRv4HM} \cite{Cotesta:2020qhw} models this mode, however, we restrict ourselves here to only $\ell=m$ modes for simplicity. Note that a higher mode model is tested against targets with only $\ell=m$ modes up to $\ell$=5.} Its performance against the target hybrids including higher modes is shown in Fig.~\ref{fig:hm_model} for mildly inclined systems. Similar to the 22 mode model, this higher mode model performs slightly better than the circular analogs, in particular at lower mass ends, as can be seen in Fig.~\ref{fig:hm_model}. However, matches are expected to improve with a model constructed adopting the approach of Sec.~\ref{sec:amplitude model} for each mode and/or with an improved dominant mode model. We defer such an exercise for a future work. 

Clearly, there is a lot of room for improvement in the model being presented, and that may be attempted in a future work. We would like to highlight a few caveats with the model at hand. The model is produced such that the merger time is at set at $t$=$0$. Since $t_{\rm{match}}$ for amplitude and frequency are obtained using analytical functions (obtained by performing multi-dimensional numerical fits), we find that a certain combination of system parameters can generate $t_{\rm{match}}>0$, implying that the inspiral and merger-ringdown waveforms are joined at a time beyond merger time. Clearly such $t_{\rm{match}}$ values cannot be used for generating the model for such realisations and should be dropped from analyses. To get a sense of the fraction of sources for which this can be expected, we generated a set of 10,000 points by randomly sampling values of a reference eccentricity ($e_0$), mass ratio ($q$) and reference mean anomaly ($l_0$) in the range\footnote{This is precisely the range from which we sample parameter values for testing the model against waveforms from the ENIGMA family.} $0.1\lesssim e_0\lesssim0.2$, $1\lesssim q\lesssim3$, and $-\pi\leq l_0\leq\pi$, respectively, and find that nearly 10\%-15\% of the points produce waveforms with $t_{\rm match}>0$.
Another caveat of our model is that it is about 25\% slower than the circular EOB model \textsc{SEOBNRv4}~\cite{PhysRevD.95.044028}. This is not surprising, as inspiral (\textsc{EccentricTD}) and merger-ringdown (\textsc{SEOBNRv4}) pieces are generated independently using \texttt{LALSuite} \cite{lalsuite}) before they are combined to obtain the model which naturally slows the construction. Note, however, that the model(s) presented in Sec.~\ref{sec:22-mode-model} and in  Appendix~\ref{sec:hm-model} can be extended to arbitrarily low frequencies since they call analytical inspiral waveforms, and calibration to target hybrids only help in identifying the transition time for inspiral and merger-ringdown attachment. It is this feature of the current approach that makes these waveforms useful. Compared to the waveforms developed in Fourier space, say, for instance, IMRPhenomXAS~\cite{Pratten:2020fqn}, these waveforms (like \textsc{SEOBNRv4}~\cite{PhysRevD.95.044028}) are much slower (by a factor of $\sim100$). 

\section{Acknowledgments}
We thank Prayush Kumar for useful comments on the manuscript. We thank K. G. Arun and Guillaume Faye for useful discussions and Sumit Kumar for help with the parameter estimation pipeline (\texttt{PyCBC Inference}). 
We thank the authors of Ref.~\cite{Chen:2020lzc} for providing a code implementation of the ENIGMA waveform model, and Prayush Kumar for technical discussions about the use of ENIGMA. We are thankful to the SXS Collaboration for making a public catalog of numerical relativity waveforms and, in particular, Ian Hinder and Harald Pfeiffer for clarifying our doubts at various stages of the project. T.R.C. thanks the members of the gravitational wave group at the Department of Physics, IIT Madras for organizing the weekly journal club sessions and the insightful discussions. A. G. acknowledges support, in part, by the Navajbai Ratan Tata Trust and the LIGO-India project at IUCAA, India. The authors are grateful for computational resources provided by the LIGO Laboratory and supported by National Science Foundation Grants No. PHY-0757058 and No. PHY-0823459. The authors also acknowledge the workstation ``powehi" in the Department of Physics, IIT Madras that was used to perform most of the parameter estimation runs shown in this paper. This document has LIGO preprint number LIGO-P2200106.
\appendix

\section{Details of Parameter Estimation Study}
\label{sec:pe-details}

\begin{figure*}[hp!]
      \centering
      \includegraphics[trim=10 10 110 120, clip, width=\linewidth]{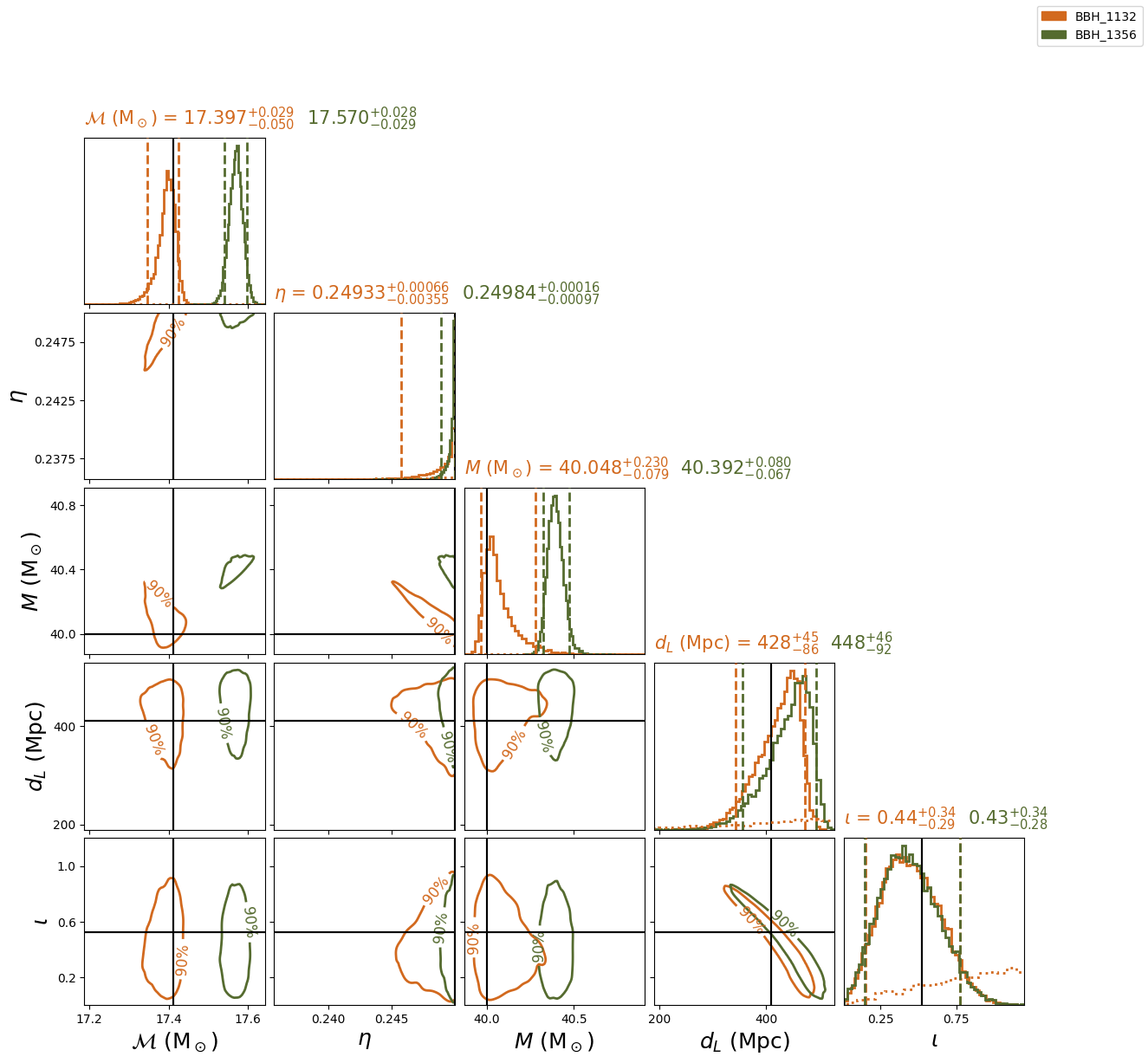}
      \caption{Corner plot for $q$=$1$, where circular (SXS:BBH:1132) simulation is shown in orange, and eccentric (HYB:SXS:BBH:1356) simulation is shown in dark green. Histograms on the diagonal show marginalized 1D posteriors, whereas the contours denote the joint 2D posteriors for various parameters. The vertical dashed lines in 1D histograms mark 90\% credible intervals, and dotted lines in orange show the prior. The black lines mark the injected values for various system parameters.}
    \label{fig:corner_q_1}
  \end{figure*}
  
\begin{figure*}
      \centering
      \includegraphics[trim=10 10 110 120, clip, width=\linewidth]{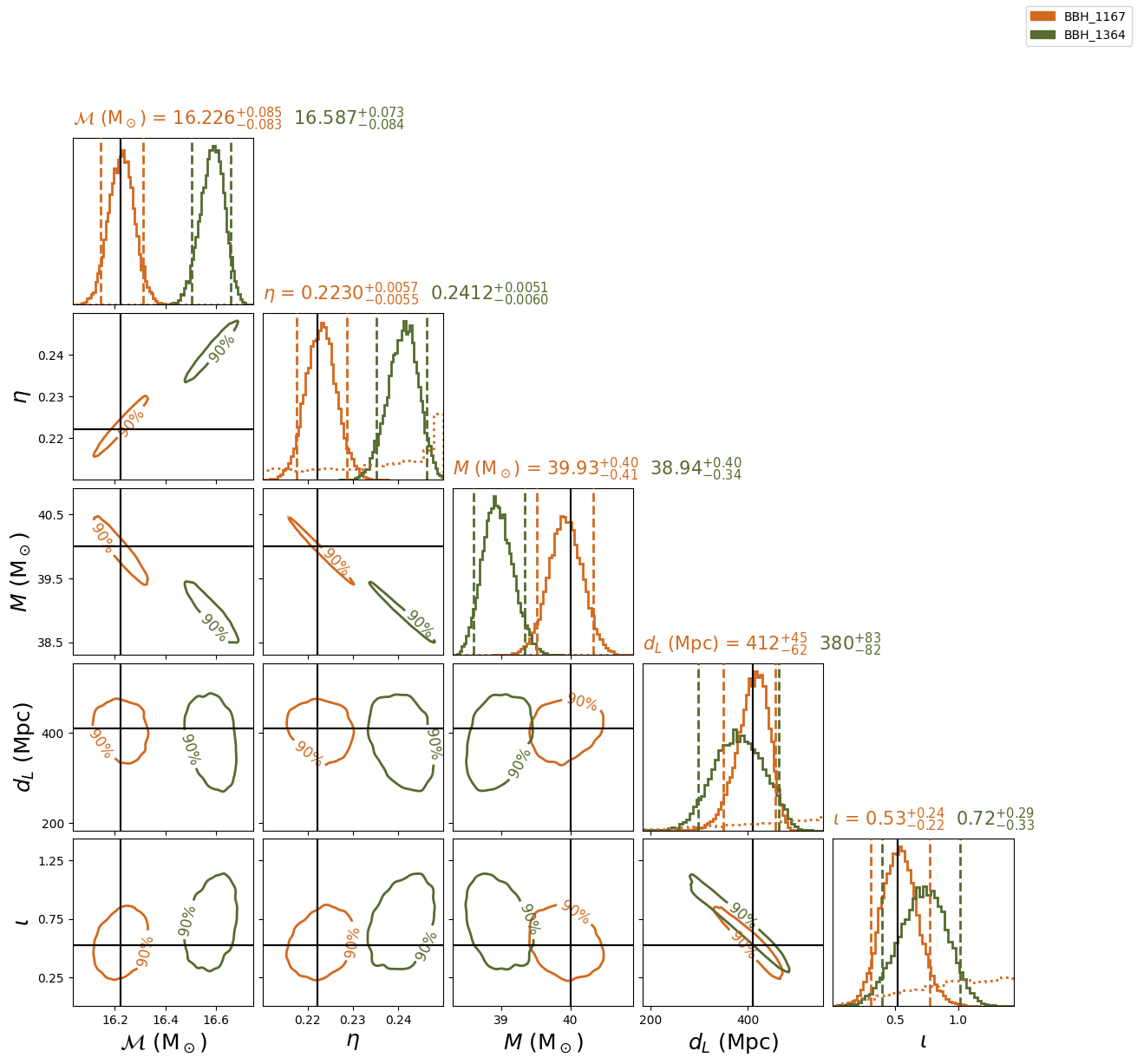}
      \caption{Corner plot for $q$=$2$, where circular (HYB:SXS:BBH:1167) simulation is shown in orange, and eccentric (HYB:SXS:BBH:1364) simulation is shown in dark green. Histograms on the diagonal show marginalized 1D posteriors, whereas the contours denote the joint 2D posteriors for various parameters. The vertical dashed lines in 1D histograms mark 90\% credible intervals, and dotted lines in orange show the prior. The black lines mark the injected values for various system parameters.}
    \label{fig:corner_q_2}
  \end{figure*}
  
\begin{figure*}
      \centering
      \includegraphics[trim=10 10 110 120, clip, width=\linewidth]{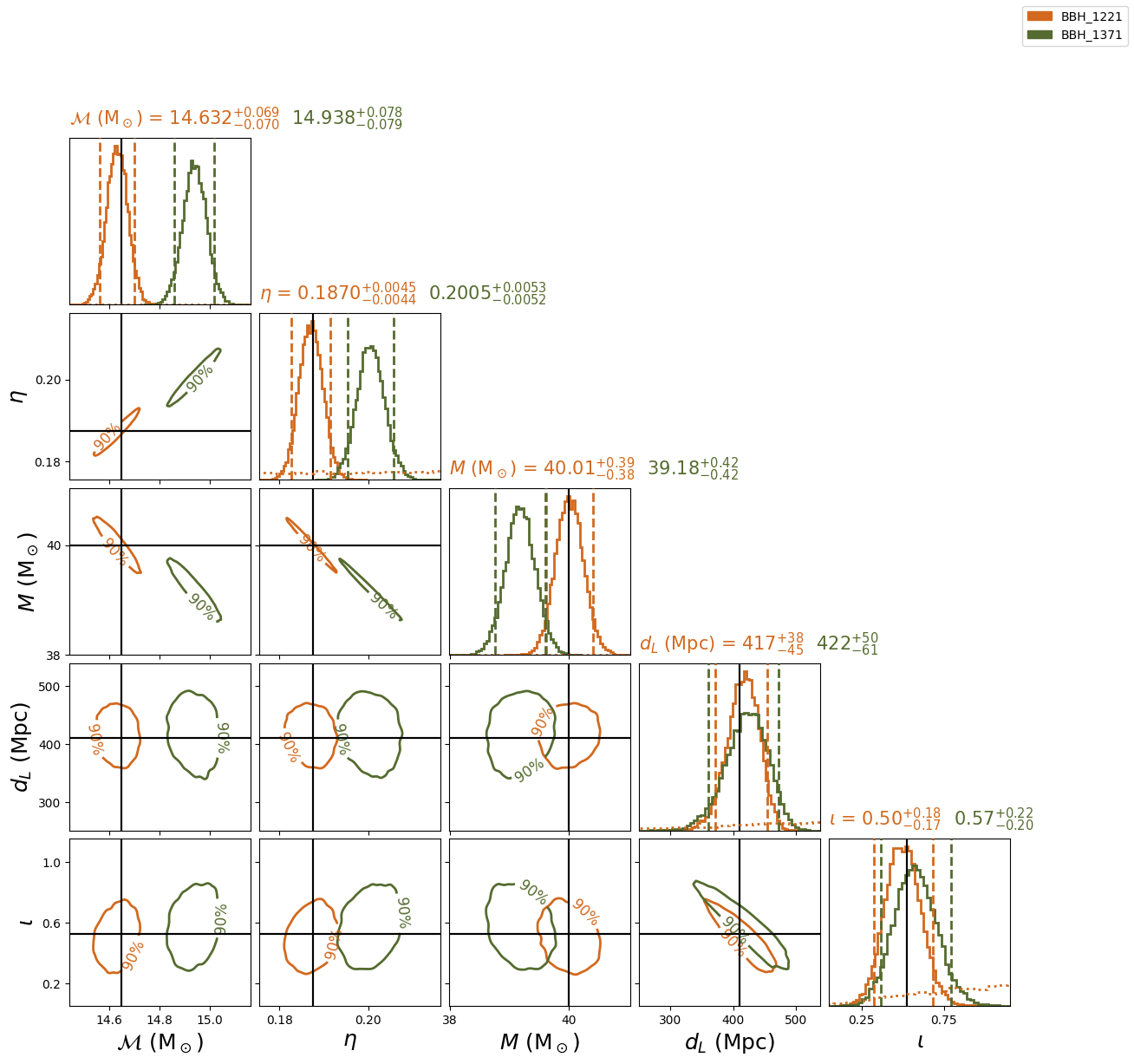}
      \caption{Corner plot for $q$=$3$, where circular (SXS:BBH:1221) simulation is shown in orange, and eccentric (HYB:SXS:BBH:1371) simulation is shown in dark green. Histograms on the diagonal show marginalized 1D posteriors, whereas the contours denote the joint 2D posteriors for various parameters. The vertical dashed lines in 1D histograms mark 90\% credible intervals, and dotted lines in orange show the prior. The black lines mark the injected values for various system parameters.}
    \label{fig:corner_q_3}
  \end{figure*}

In this section, we include the corner plots for all the injections that have been performed in order to highlight the correlations between various binary parameters. The parameters chosen for corner plots are chirp mass ($\mathcal{M}_c$), total mass ($M$), symmetric mass ratio ($\eta$), luminosity distance ($d_L$), and the binary's inclination angle with respect to the line of sight ($\iota$). Figures \ref{fig:corner_q_1}-\ref{fig:corner_q_3} show the corner plots for $q=1, 2,$ and $3$, respectively (see also~\cite{OShea:2021ugg}). These plots clearly show the correlations among various mass parameters ($\mathcal{M}_c$, $M$, and $\eta$) and strong correlations between luminosity distance $d_L$ and inclination angle ($\iota$). \\
\\
It is interesting to note that the symmetric mass ratio $\eta$ has an upper limit of $\eta\leq0.25$ (by definition), thus for $q$=$1$, since $\eta=0.25$, the posterior hits the prior boundary on the right. Now, because of the correlations between chirp mass ($\mathcal{M}_c$) and $\eta$, this translates into slight shifting of the posterior which results in the injection value not coinciding with the median of the posterior. This is observed both in Figs.~\ref{fig:pe_systematics} and \ref{fig:corner_q_1}. This feature also appears when we inject a signal using \textsc{IMRPhenomXHM} and recover using the same waveform, thus indicating that this is solely because of the correlations between $\eta$ and $\mathcal{M}_c$ and not because of differences in the injected and recovery waveform. The slight shift in $q$=$3$ posterior of circular injection (HYB:SXS:BBH:1221) though can be attributed to the slight differences in the injected and recovered signal. \\
\\
Another interesting feature of these plots is that, while the mass parameters show a shift in the posterior with eccentric injections, the distance and inclination posteriors are still able to recover the injected values, even for eccentric injections. This hints at the fact that, while the eccentricity parameter is strongly correlated with the mass parameters of the binary, it is not so with the case of extrinsic parameters like luminosity distance and inclination angle (see also~\cite{OShea:2021ugg}).

\clearpage

\section{Extending to a higher modes model}
\label{sec:hm-model}

\begin{figure*}[htbp!]
\centering
    \includegraphics[width=0.495\textwidth]{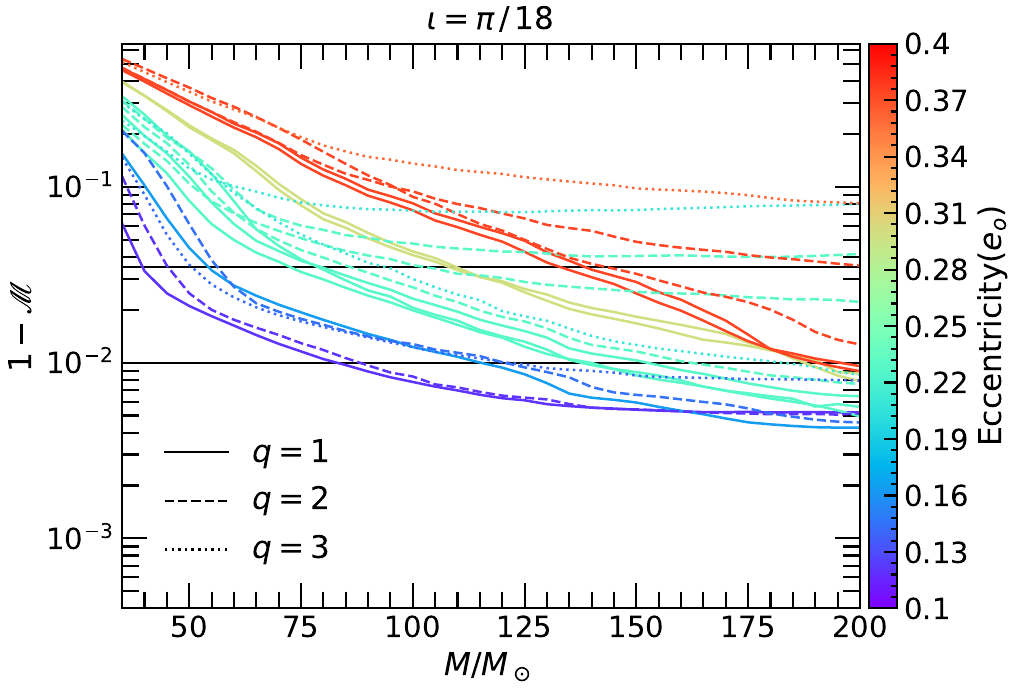}
    \includegraphics[width=0.495\textwidth]{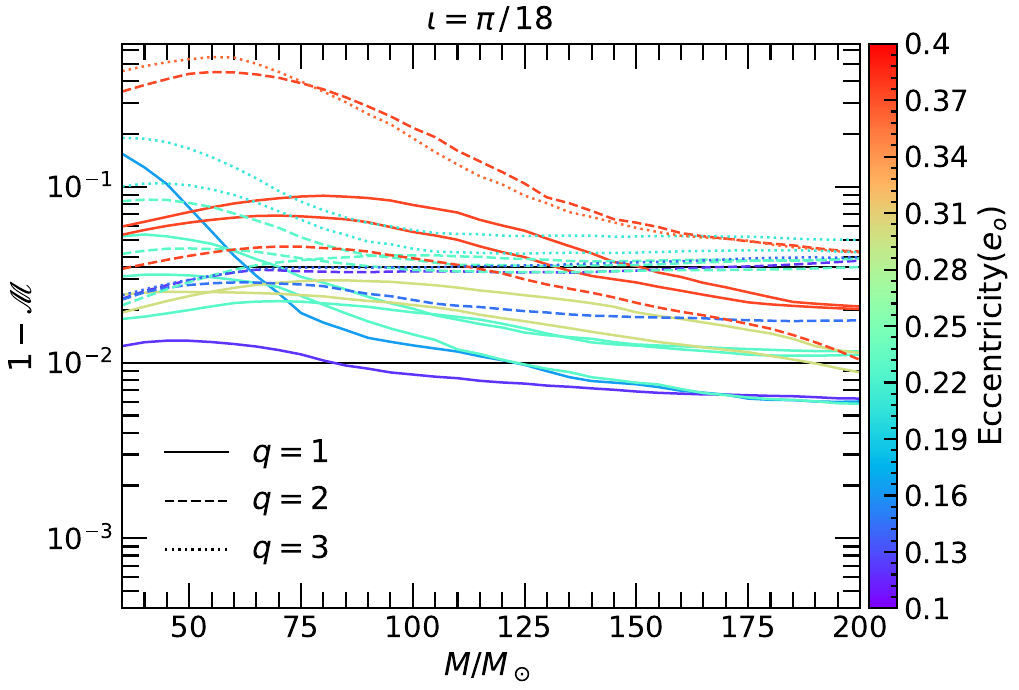}
    \includegraphics[width=0.495\textwidth]{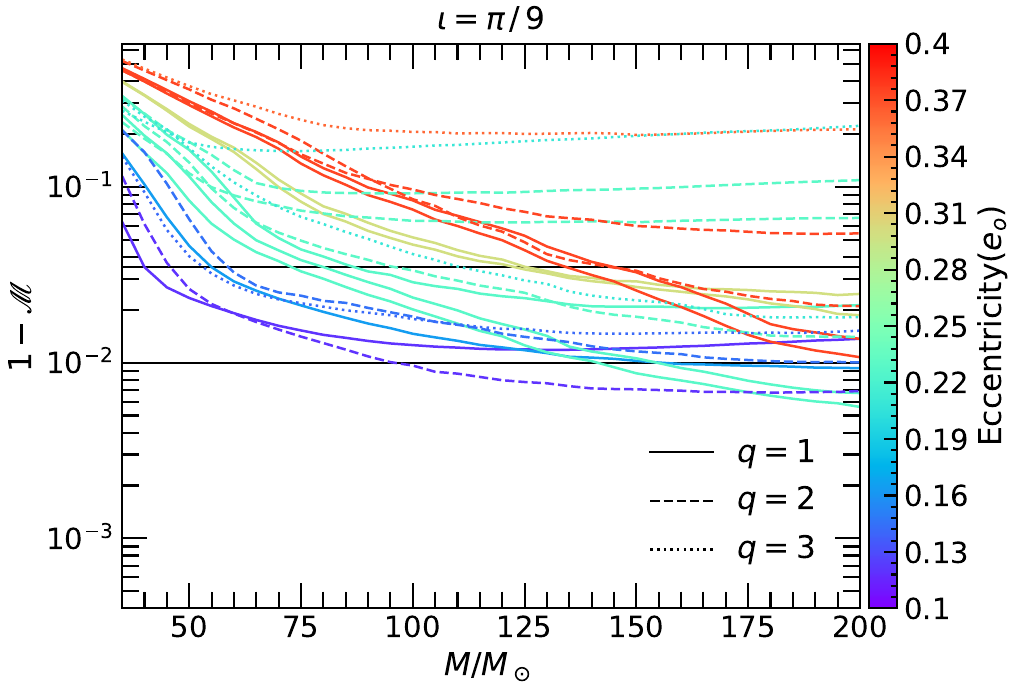}
    \includegraphics[width=0.495\textwidth]{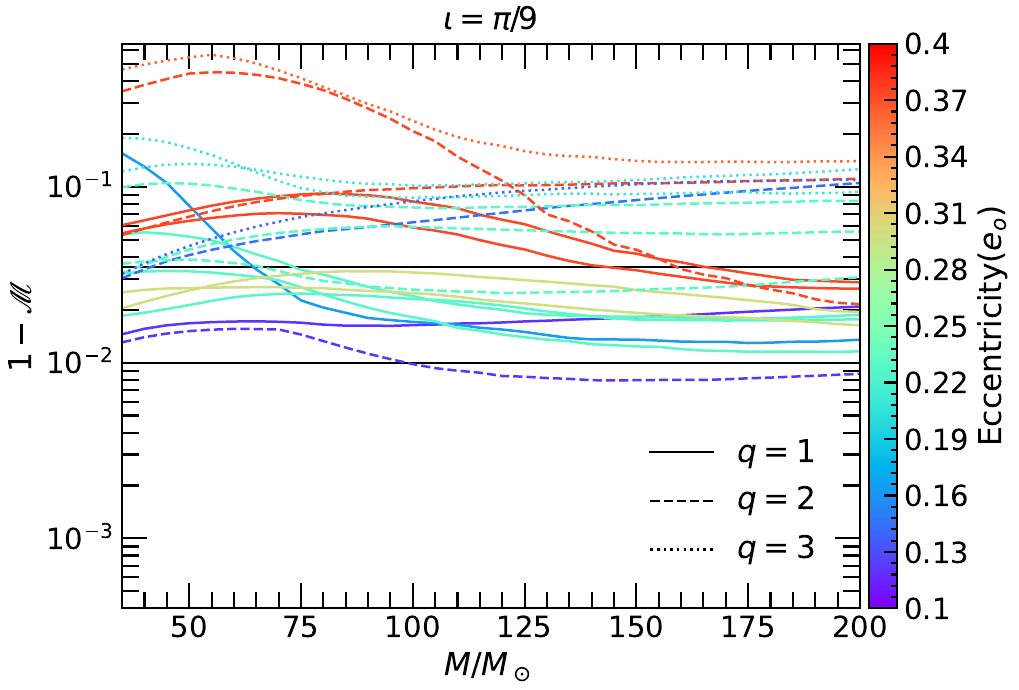}
    \includegraphics[width=0.495\textwidth]{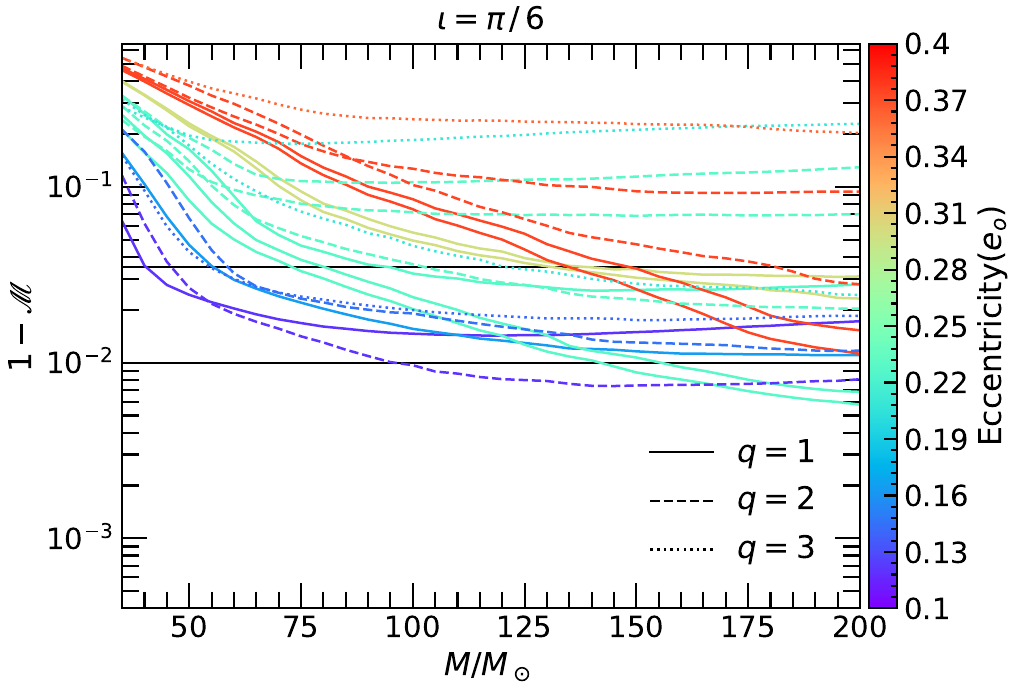}
    \includegraphics[width=0.495\textwidth]{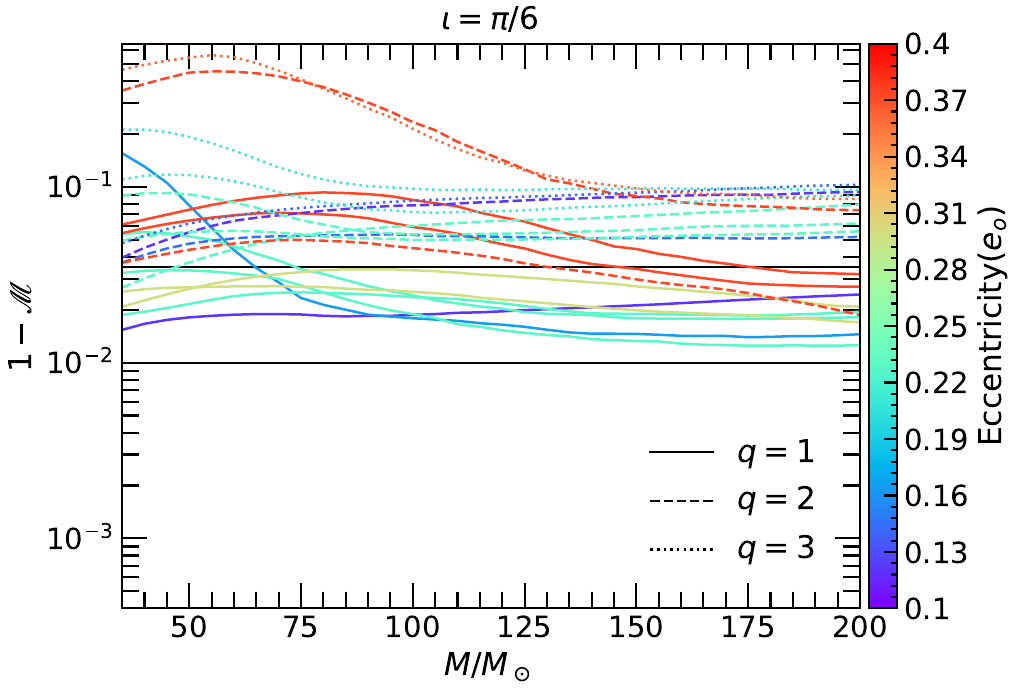}
   
    \caption{Mismatches with a set of twenty $(2, \pm2)$, $(3, \pm3)$, $(4, \pm4)$, and $(5, \pm5)$ mode hybrids with a higher mode quasi-circular waveform \textsc{SEOBNRv4HM} \cite{Cotesta:2020qhw}  (left) and with the higher mode model based on the dominant mode prescription of Sec.~\ref{sec:22-mode-model} (right), including the exact same set of modes for three mildly inclined systems ($\iota$=$\pi/18, \pi/9, \pi/6$) shown in top, middle, and bottom panels, respectively. The two horizontal lines report 96.5\% and 99\% agreement, respectively. The eccentricity values displayed with color bars are computed at $x_0 = 0.045$ for all hybrids.}
    \label{fig:hm_model}
\end{figure*}

Here we discuss the performance of a simple higher mode model obtained using the prescription for the dominant mode obtained in Sec.~\ref{sec:22-mode-model} for all $\ell=m$ modes up to $\ell=5$ i.e. the model includes ($\ell$, $|m|$)=(2, 2), (3, 3), (4, 4), and (5, 5) modes. [Note that, in principle, we could also model the (2, 1) mode as \textsc{SEOBNRv4HM} \cite{Cotesta:2020qhw} which constitutes the merger-ringdown piece models this mode.  However, we restrict ourselves here to only $\ell=m$ modes for simplicity.] We simply use the analytical expressions for $t_{\rm{shift}}$ and $t_{\rm{match}}$, obtained for the dominant mode model in Sec.~\ref{sec:22-mode-model}, to each non-quadrupole mode in order to join the inspiral and merger-ringdown pieces. The inspiral piece of this  higher mode model is obtained by combining the 3PN accurate amplitude expressions from Refs.~\cite{Boetzel:2019nfw, Ebersold:2019kdc} (also used in constructing the hybrids presented in Table~\ref{tab:amplitude}) with the phasing prescription of Ref. \cite{Tanay:2016zog}. On the other hand, the merger-ringdown waveform is the quasi-circular waveform \textsc{SEOBNRv4HM} \cite{Cotesta:2020qhw}, which models ($\ell$, $|m|$)=(2, 2), (2, 1), (3, 3), (4, 4), and (5, 5) modes. Mismatch plots of Fig. \ref{fig:hm_model} show the performance of the model compared to \textsc{SEOBNRv4HM} \cite{Cotesta:2020qhw} for three mildly inclined systems (10$^{\circ}$, 20$^{\circ}$, 30$^{\circ}$). The left panels show mismatches with quasi-circular model \textsc{SEOBNRv4HM} \cite{Cotesta:2020qhw} while the right panels display mismatches with the higher mode model obtained here. Note that both the target (hybrids) and the templates (higher mode model or the quasi-circular \textsc{SEOBNRv4HM}) include ($\ell$, $|m|$)=(2, 2), (3, 3), (4, 4), and (5, 5) modes. Similar to the 22 mode model, this higher mode model performs slightly better at high mass ends compared to the quasi-circular model, but much better at lower mass ends as can be seen in Fig.~\ref{fig:hm_model}. For instance, mismatches are typically smaller than $5\%$(20\%) for systems heavier (lighter) than $100\,M_{\odot}$ except for a few high eccentricity cases. Mismatches with quasi-circular templates, on the other hand, are typically smaller than $10\%$(60\%) for systems heavier (lighter) than $100\,M_{\odot}$. Assuming the dominant mode model prescriptions for attaching the inspiral and merger-ringdown pieces may not be optimal, one may hope to improve these by extending the methods discussed in Sec.~\ref{sec:22-mode-model} to obtain independent prescriptions for each mode.

\appendix

\begin{table*}[h]
        \centering
        \begin{tabular}{c | c c c c c}
            \hline
            $A_{\alpha \beta 0 0}$ & $\beta = 0$ & 1 & 2  \\
\hline
$\alpha = 0$ & $-1198.5024$   & $-700.73289$  & $1483.6558$   \\
1            & $45323.781$    & $-25867.453$  & $-11881.667$  \\
2            & $-277961.63$   & $205622.54$   & $19756.350$   \\
3            & $480903.69$    & $-394215.78$  & $0$           

        \end{tabular}
        
        \vspace*{0.5cm}
        
        \centering 
        \begin{tabular}{c|c|cc}
            \hline
            $\alpha \beta \gamma \delta$ & $A_{\alpha \beta \gamma \delta}$ & $a_{\alpha \beta \gamma \delta}$\\
\hline
$1110$ & $-1781.3755$    & $23.049324$  \\
$2201$ & $8601.0938$     & $12.644587$  \\
$3110$ & $24719.217$     & $35.365563$  \\
$1210$ & $218.37381$     & $-2.5655030$ 

        \end{tabular}

        \caption{Table of coefficients for the analytical expression of $t_{\rm{shift}}$ in Eq.~\eqref{eq:t_shift_analytical}. All other coefficients not included in the table are zero. }
        \label{tab:shift}
\end{table*}

\begin{table*}[h]
        \centering
        \begin{tabular}{c | c c c c c}
            \hline
            $B_{\alpha \beta 0 0}$ & $\beta = 0$ & 1 & 2  \\
\hline
$\alpha = 0$ & $317.56597$              & $107515.46$             & $-449629.32$              \\
1            & $-72361.787$             & $-913714.19$            & $4.9085757 \times 10^6$   \\
2            & $889870.52$              & $-953568.19$            & $-1.0660253 \times 10^7$  \\
3            & $-2.0028862 \times 10^6$ & $8.1132219 \times 10^6$ & $0$                       

        \end{tabular}

        \vspace*{0.5cm}
        
        \centering 
        \begin{tabular}{c | c | c c}
            \hline
            $\alpha \beta \gamma \delta$ & $B_{\alpha \beta \gamma \delta}$ & $b_{\alpha \beta \gamma \delta}$\\
\hline
$1110$ & $8589.8002$    & $220.81235$  \\
$2210$ & $-155537.19$   & $-1123.6840$ \\
$3101$ & $57429.241$    & $2257.6694$  

        \end{tabular}

        \caption{Table of coefficients for the analytical expression of amplitude $t_{\rm{match}}$ in Eq.~\eqref{eq:t_match_amplitude}. All other coefficients not included in the table are zero. }
        \label{tab:amplitude}
\end{table*}

\begin{table*}[h]
        \centering
        \begin{tabular}{c | c c c c c}
            \hline
            $C_{\alpha \beta 0 0}$ & $\beta = 0$ & 1 & 2  \\
\hline
$\alpha = 0$ & $374015.60$                & $-3.9854127 \times 10^6$   & $1.0318033 \times 10^7$   \\
1            & $-1.5126993 \times 10^6$   & $1.8270088 \times 10^7$    & $-5.4439513 \times 10^7$  \\
2            & $1.0112138 \times 10^6$    & $-1.6624831 \times 10^7$   & $7.1280945 \times 10^7$   \\
3            & $2.6114991 \times 10^6$    & $-1.7913253 \times 10^7$   & $0$                       
        \end{tabular}

        \vspace*{0.5cm}
        
        \centering 
        \begin{tabular}{c|c|cc}
            \hline
            $\alpha \beta \gamma \delta$ & $C_{\alpha \beta \gamma \delta}$ & $c_{\alpha \beta \gamma \delta}$\\
\hline
$1110$ & $-317485.702$             & $123.07063$ \\
$2210$ & $5.7669647 \times 10^6$   & $883.39202$ \\
$3101$ & $1.0347922 \times 10^7$   & $3823.0109$ \\
$3201$ & $5.2523026 \times 10^7$   & $16116.188$ 
        \end{tabular}

        \caption{Table of coefficients for the analytical expression of frequency $t_{\rm{match}}$ in Eq.~\eqref{eq:t_match_frequency}. All other coefficients not included in the table are zero. }
        \label{tab:freq}
\end{table*}

\clearpage

\bibliographystyle{apsrev4-1}
\bibliography{EccPNNRHigherModes}
\end{document}